\documentclass[pre,aps,showpacs,twocolumn]{revtex4}
\usepackage{amsmath,amssymb}
\usepackage{graphicx}
\providecommand{\select}[2]{\genfrac{\{}{\}}{0pt}{}{#1}{#2}}
\providecommand{\dd}{\mbox{d}}
\providecommand{\dt}{\frac{\partial}{\partial t}}
\providecommand{\nn}{\nonumber}
\begin{document}
\title{Shear induced instabilities
        in layered liquids}
\date{\today}

\author{
        {G\"unter K. Auernhammer}}
\email{guenter.auernhammer@uni-bayreuth.de}
\author{
        {Helmut R. Brand}}
\affiliation{Theoretische Physik III, Universit\"at Bayreuth, 95440
  Bayreuth, Germany} 
\author{
        {Harald Pleiner}}
\affiliation{Max-Planck-Institute for Polymer Research, POBox 3148,
  55021 Mainz, Germany} 

\begin{abstract}
  Motivated by the experimentally observed shear-induced
  destabilization and reorientation of smectic $A$ like systems, 
  we consider an extended formulation of smectic $A$ hydrodynamics.
  We
  include both, the smectic layering (via the layer displacement $u$
  and the layer normal $\hat p$) and the director $\hat n$ of the
  underlying nematic order in our macroscopic hydrodynamic description
  and allow both directions to differ in
  non equilibrium situations. In an homeotropically aligned sample the 
  nematic director does couple to an applied simple shear, whereas the
  smectic layering stays unchanged. This difference leads to a finite
  (but usually small) angle between $\hat n$ and $\hat p$, which we
  find to be equivalent to an effective dilatation of the
  layers. This effective dilatation leads, above a certain threshold,
  to an undulation instability of the layers. We generalize our earlier
  approach [Rheol. Acta {\bf 39} (3), 15] 
  and include the cross couplings with the velocity field
  and the order parameters for orientational and positional order
  and show how the order parameters interact with the undulation
  instability. 
  We explore the influence of various material parameters on the
  instability. 
  Comparing our results to recent experiments and molecular dynamic
  simulations, we find a good qualitative agreement.
\end{abstract}
	
\maketitle

\section{Introduction}
Submitted to an applied shear flow, many complex liquids show an
interesting coupling between their internal structure and the flow
field. 
For smectic $A$ like systems (including block copolymers, lyotropic
systems and side-chain liquid crystalline polymers) this coupling may
induce reorientation of the layers.
Experiments on a variety of systems which differ significantly in
their microscopic  
details show nevertheless striking similarities in their macroscopic
behavior under shear. The systems under investigation include 
block copolymers
\cite{gupta96,wiesner97,laurer99,zryd98,leist99,polis99},
low molecular weight (LMW) liquid crystals
\cite{horn78,safinya91,panizza95}, 
lyotropic lamellar phases (both LMW
\cite{diat93,mueller99,escalante00} and polymeric \cite{zipfel99b}),
and liquid crystalline side-chain polymers \cite{noirez97,noirez00}.
These experiments use either a steady shear (typically for the low
viscosity systems e.g. in a Couette cell) or large amplitude
oscillatory shear (often in the highly viscous polymeric systems,
e.g. in a cone-plate or plate-plate geometry). Due to these
experimental differences a direct comparison between the different
systems is not always straightforward.
The common features of all these experiments can be described as
follows. 
Starting with
an aligned sample where the layers are parallel to the planes of
constant velocity (``parallel'' orientation), the
layering is stable up to a certain 
critical shear rate
\cite{safinya91,panizza95,diat93,mueller99,zipfel99b,wiesner97,leist99}.
At higher shear rates two different situations
are observed. Depending on the system either multi-lamellar vesicles
\cite{diat93,escalante00,zipfel99b} (``onions'', typically in
lyotropic systems) or layers perpendicular to the
vorticity direction
\cite{safinya91,panizza95,gupta96,wiesner97,laurer99,zryd98,leist99,noirez97} 
(``perpendicular'' orientation, typically in solvent free systems)
form. In  
some of the systems a third regime is observed at even higher shear
rates with a parallel orientation \cite{diat93,leist99}. 
If the starting point is rather a randomly distributed lamellar phase, 
the first regime is not observed \cite{gupta96,zryd98,wang99,escalante00}. 
This last point illustrates that experiments on layered liquids depend 
on the history of the sample. In our further discussion we will
restrict ourselves to systems showing a well aligned parallel
orientation before shear is applied.

The experimental similarities between different systems indicate, that
the theoretical description of these
reorientations can be 
constructed, at least to some extent, from a common basis independent
of the actual system (on the other hand, a 
description including the differences between the systems under
investigation must refer closer to their microscopic details). When
looking for a macroscopic description, 
the well established standard smectic $A$ hydrodynamics
\cite{degennes72,martin72,degennes93,pleiner96}
is a good starting point for such a theoretical
approach. 

As first shown by Delaye et al. \cite{delaye73} and Clark and Meyer
\cite{clark73} thermotropic smectic $A$ liquid crystals are very
sensitive against dilatations of the layers. Above a very small, but
finite, critical dilatation the liquid crystal develops undulations of
the layers to reduce the 
strain locally. Oswald and Ben-Abraham considered dilated smectic $A$
under shear \cite{oswald82b}.
When a shear flow is applied (with a parallel
orientation of the layers), the onset for  undulations is
unchanged only if the wave vector of the undulations points in the
vorticity direction (a similar situation was later considered in
\cite{wunenburger00}). Whenever this wave vector has a component in 
the flow direction, the onset of the undulation instability is
augmented by a portion proportional to the applied shear rate. 
No destabilizing  mechanism for well aligned parallel layers is
present in the standard smectic $A$ hydrodynamics. 

Recently we
proposed an extended hydrodynamic description
\cite{auernhammer00,auernhammer00a} of smectic $A$ liquid
crystals. Using 
both, the director of the underlying nematic order and the layer
normal of the smectic layers, we showed the possibility of a
shear-induced undulation instability in well
aligned parallel layers. Within the framework of irreversible
thermodynamics (which allows the inclusion of dissipative as well as
reversible effects)
we derived macroscopic hydrodynamic equations for the
system and performed a linear stability analysis of these equations
(using a number of approximations). 
As always, a linear stability analysis is limited
to the onset of the first instability. 
Other theoretical approaches to these reorientation phenomena have been
undertaken by Bruinsma and Rabin \cite{bruinsma92}, Zilman and Granek
\cite{zilman99} (both papers are considering the influence of the
shear on layer fluctuations) and Williams and MacKintosh
\cite{williams94} (minimizing a free energy density including a
coupling to the applied shear stress). To our knowledge, no macroscopic
hydrodynamic approach besides \cite{auernhammer00,auernhammer00a} has
been published up to now. 

The present paper is structured as follows: After a brief review of
the model in Sec. \ref{sec:mod} and its implementation in
Sec. \ref{sec:imp} we extend the basic model of 
\cite{auernhammer00,auernhammer00a} in the following sections.
Especially we include the cross coupling to the velocity field and
the moduli of the nematic and smectic order parameters. It turns out
that the coupling terms to the velocity are important since they can
change the 
critical parameters significantly. We find that the moduli of the
order parameters also show undulations and, thus, regions with a
reduced order parameter can be identified.  
The comparison of the different levels of approximations
shows that the basic model is contained in this more general analysis
as a special case. We also compare our results to experiments and
molecular dynamic simulations and show that an oscillatory instability
is extremely unlikely to occur.  

\section{Model and technique}
\label{sec:mod-tec}
\subsection{Physical idea of the model} 
\label{sec:mod}
\begin{figure}
  \caption{\label{fig:geometry}
    At the level of the approximation we use in this paper, all
    experimental shear geometries are equivalent to a simple steady
    shear. We choose our system of coordinates such that the normal to
    the plates points along the $z$-axis and the plates are
    located at $z = \pm \frac d2$. 
    }
  \begin{center}
    \includegraphics*[height=\columnwidth, angle=-90]{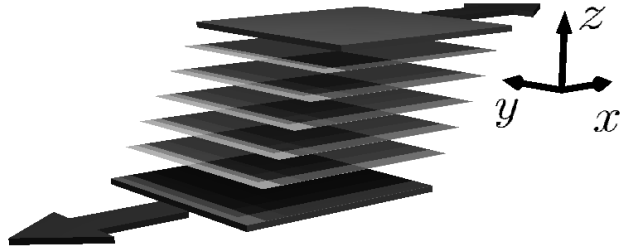}
  \end{center}
\end{figure}
In a smectic $A$ liquid crystal one can easily define two directions: the
normal to the layers $\hat p$ and an average over the molecular axes,
the director, $\hat n$. In the standard formulation of smectic $A$
hydrodynamics these two directions are parallel by construction. Only
in the vicinity of phase transitions (either the nematic--smectic $A$ 
or smectic $A$--smectic $C^*$) it has been
shown that director fluctuations are of physical interest
\cite{degennes73,litster79,garoff77}. 
Nevertheless $\hat n$ and $\hat p$ differ 
significantly in their interaction with an applied shear flow.

We consider a situation as show in Fig.~\ref{fig:geometry}. A well
aligned smectic $A$ liquid crystal is placed between two parallel and
laterally infinite plates. The upper plate (located at $z = d/2$)
moves with a constant velocity $\vec v_u = d \dot \gamma \hat e_x/2$
to the right and the lower plate (at $- d/2$) moves 
with the same velocity in the opposite direction ($\vec v_l = -
d \dot \gamma \hat e_x/2$). Thus the sample is submitted to an 
average shear given by $(v_u - v_l)/d = \dot \gamma$. 
As mentioned above, a three dimensional stack of parallel fluid layers
cannot couple directly to an applied shear flow. 
Neither does the layer normal: it stays unchanged as long as the flow
direction lies within the layers. In contrast, it is well
known from nematodynamics that the director experiences a torque in a
shear flow. This torque leads --- in the simplest case --- to a flow
aligning behavior of the director. The key assumption in the model of 
\cite{auernhammer00} is that this torque is still present in a smectic
$A$ liquid crystal and acts only on the director $\hat n$ and not on
the layer normal $\hat p$. An energetic coupling between $\hat n$ and
$\hat p$ ensures that both directions are parallel in equilibrium. 

\begin{figure}
  \caption{\label{fig:effdilat}
    A finite angle $\theta$ between $\hat n$ and $\hat p$ leads to a 
    tendency of the layers to reduce their thickness. Supposing a
    constant number of layers in the sample, this tendency is
    equivalent to an effective dilatation of the layers. For small
    angles between $\hat n$ and $\hat p$ the relative effective
    dilation is given by $\theta^2/2$.
    }
  \begin{center}
    \includegraphics*[height=\columnwidth, angle=-90]{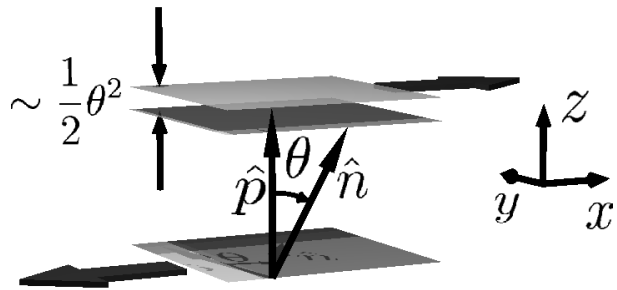}
  \end{center}
\end{figure}
\begin{figure}
  \caption{\label{fig:undul}
    Above a certain threshold the effective dilatation due to the
    director tilt will lead to buckling of the layers. Note the
    difference in directions: the director is tilted in the flow
    direction, whereas the wave vector points along the $y$-axis. This
    configuration cancels the direct coupling between the flow and the
    buckling. 
}
  \begin{center}
    \includegraphics*[height=\columnwidth, angle=-90]{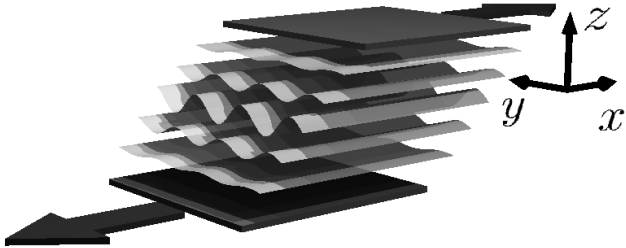}
  \end{center}
\end{figure}

Submitted to a shear flow the layer normal will stay unchanged, but
the director will tilt in the direction of the flow until the torques
due to the flow and due to the coupling to the layer normal balance one
another. For any given shear rate a finite, but usually small, angle
$\theta$ between $\hat n$ and $\hat p$ will result. This finite angle
has important consequences for  
the layers: Since the preferred thickness of the layers is proportional
to the projection of the director on the layer normal, a finite angle
between those two directions is equivalent to an effective
dilatation of approximately $\theta^2/2$ (see
Fig.~\ref{fig:effdilat}). If we assume a constant total sample
thickness and exclude effects of defects, the system can accommodate
this constraint by layer rotations. A global rotation of the layers is
not possible, but they can rotate locally (as in the case of
dilated smectic $A$ liquid crystals \cite{clark73,delaye73}). This
local rotation of the layers leads to undulations as shown in
Fig.~\ref{fig:undul}. 
These undulation are a compromise between the
effective dilatation (which is not favorable for the system) and the
curvature of the layers due to the undulations (which costs energy). 
In the static case of dilated smectic $A$ liquid crystals no direction
is preferred, but Oswald and Ben-Abraham \cite{oswald82b} have shown
that this symmetry is broken if an additional shear is applied to the
system. In this case the standard formulation of smectic $A$
hydrodynamics predicts that the wave vector of the undulations will
point along the neutral direction of the shear. In this paper we will
assume that this result of Oswald and Ben-Abraham also holds in the
case of our extended formulation of smectic $A$ hydrodynamics
(see Fig.~\ref{fig:undul}).

\subsection{Implementation of the model}
\label{sec:imp}
To generate the macroscopic hydrodynamic equations we follow
the procedure given by the framework of irreversible thermodynamics
\cite{degroot62}. This method has 
successfully been applied in many cases to derive the macroscopic
hydrodynamic equations of complex fluids (see
e.g. \cite{auernhammer00,martin72,liu79,pleiner96}).
The advantage of this method is its systematic way of deducing the
governing equations. Once the set of variables is given, the
macroscopic hydrodynamic equations follow by applying basic symmetry
arguments and thermodynamic considerations. 

Let us briefly review the essential ingredients to this procedure
(for more details of the method see \cite{pleiner96} and for our model
\cite{auernhammer00}). 
For a given system the hydrodynamic variables can be split up in two
categories: variables reflecting conserved quantities (e.g. the linear
momentum density, the mass density etc.) and variables due to
spontaneously broken continuous symmetries (e.g. the nematic director or 
the layer displacements of the smectic layers). Additionally, in some
cases non-hydrodynamic variables (e.g. the strength of the order
parameter  \cite{brand86a}) 
can show slow dynamics which can be described within this
framework (see e.g. \cite{liu79,pleiner96}). 

Using all these variables the relations, which form the starting point
for the further calculations, can be constructed. 
These relations are: the energy density
$\epsilon$, the dissipation 
function $R$, the \mbox{Gibbs-,} and the Gibbs-Duhem relation.
To illustrate the idea of our model we split up $\epsilon$ and $R$
into 
several parts according to the different origin of the variables. 
\begin{itemize}
\item[-] conserved quantities (index $cons$)
\item[-] symmetry variables (index $sym$)
\item[-] the modulus of the order parameter (index $ord$) 
\end{itemize}
In the spirit of our model two order parameters play a role: the
nematic tensorial order parameter $Q_{ij}$ and the smectic $A$ complex 
order parameter $\Phi$. 
For practical reason we use the director $\hat 
n$ and the modulus $S^{(n)}_{\mbox{}}$ in the uniaxial nematic case
[$Q_{ij} = \frac 32 S^{(n)}_{\mbox{}}(n_i n_j - \frac 13
\delta_{ij})$] and the layer displacement $u$ and the modulus 
$S^{(s)}_{\mbox{}}$ in the smectic $A$ case [$\varphi = S^{(s)}_{\mbox{}}
\exp\{iq_0(z-u)\}$].  Here, as in the rest of the paper, we refer to
the system of coordinates defined in Sect.~\ref{sec:mod}.
We note that $u$ is only a good variable if we consider small
deformations of the layers. For large layer deformations the phase
$\varphi = z - u$ is the appropriate variable
\cite{marignan83,pleiner99}. 
In our further discussion, we will concentrate on the parts due to
symmetry variables and the order parameters, while for terms already
present in the isotropic fluid see e.g. \cite{pleiner96,degroot62}. 
\begin{table}[tbp]
  \begin{center}
    \begin{ruledtabular}
    \begin{tabular}[t]{lcl}
      symbol & explicit form & comment \\[1ex]
      \hline
      $K$ & $K$ & bending modulus of layers\\[1ex]
      $B_0$ & $B_0$ & compressibility of layers \\[1ex]
      $B_1$ & $B_1$ & coupling between the director \\
         &&and the layer normal\\[1ex]
      $L_0^{(n,s)}$ & $L_0^{(n,s)}$ & variations of the order \\
         && parameter (nematic and \\
         && smectic, respectively)\\[1ex]
      $L_{1,ij}^{(n)}$ & $ L_\perp^{(n)} \delta_{ij}^\perp 
        + L_\parallel^{(n)} n_i n_j$
        & gradients terms of the order \\
        && parameter (nematic) \\[1ex]
      $M_{ijk}$ 
        & $M_0(\delta_{ij}^{\perp} n_k + \delta_{ik}^{\perp} n_j)$
        & cross-coupling between the \\ 
        && director and order parameter\\
        && (nematic)\\[1ex]
      $L_{1,ij}^{(s)}$ & $ L_\perp^{(s)} (\delta_{ij} - p_i p_j) $
        & gradients terms of the order \\
        & \hspace{4ex} $+ L_\parallel^{(s)} p_i p_j$
        & parameter (smectic) \\[1ex]
    \end{tabular}
    \end{ruledtabular}
    \caption{Summary of the notation. In these definitions we use the
      transverse Kronecker symbol $\delta_{ij}^\perp = \delta_{ij} -
      n_i n_j$. Due to the thermodynamic stability of the systems the
      following combinations of constants must be positive: $B_0$,
      $B_1$, $K$, $L_0$, $L_\perp^{(n,s)}$, $L_\parallel$, and
      $M_0^2 - KL_\parallel$. For the last relation we
      used the equivalence of $K$ and $K_1$.
      }
    \label{tab:notation}
  \end{center}
\end{table}

Let us first consider the energy density.
The conventions of notation introduced by the following equations are 
summarized in Tab. \ref{tab:notation}. 
\begin{equation}
  \label{eq:ener_coll}
  \epsilon = \epsilon_{cons} + \epsilon_{sym} +
             \epsilon_{ord}^{(n)} + 
             \epsilon_{ord}^{(s)}
\end{equation}
$\epsilon_{cons}$, which is identical to the isotropic fluid, is
discussed elsewhere \cite{degroot62,pleiner96}. The symmetry part
reads: 
\begin{eqnarray}
  \label{eq:ener_den_sym_all}
   \epsilon_{sym} & = & \hspace{3ex}
       \frac 12 K_1 (\nabla \cdot \hat n)^2 
       + \frac 12 K_2 \left[\hat n\cdot(\nabla\times \hat n) \right]^2
       \nn \\ && {}
       + \frac 12 K_3 \left[\hat n\times(\nabla\times \hat n)
         \right]^2
       \nn \\ &&{}
       + \frac 12 K\left(\nabla^2_{\perp} u\right)^2
       \nn \\ &&{}
       + \frac 12 B_0 \left[\nabla_z u + (1-n_z) - 
         \frac 12 \left( \nabla_{\perp} u\right)^2 \right]^2 
       \nonumber \\
   &&{}+ \frac 12 B_1 \left(\hat n \times \hat p\right)^2 
\end{eqnarray}
In Eq. (\ref{eq:ener_den_sym_all})
the spirit of the model becomes clear. We combine
the properties of a nematic liquid crystal (the first two lines) with
these of a smectic $A$ (the third and fourth line) and couple both
parts (the last line) in such a way that $\hat n$ and $\hat p$ are
parallel in equilibrium. As already discussed earlier
\cite{auernhammer00}, $\epsilon_{sym}$ simplifies considerably by
dropping higher order terms and assuming a small angle between $\hat
n$ and $\hat p$. Splay deformations of the director are generally
considered as higher order corrections to dilatations of the smectic 
layers. Twist deformations are forbidden in standard smectic $A$
hydrodynamics and must be small as long as the angle between $\hat n$
and $\hat p$ is small.  
Additionally, the difference between the splay deformation of the
director field $K_1/2\; \left(\nabla \cdot \hat n\right)^2$  and
bending of the layers $K/2\; \left(\nabla_{\perp}^2 u \right)^2 $
is negligible. 
Consequently we combine splay and bend in a single term with a
single elastic constant which we call $K^\prime$:
$K_1/2\; \left(\nabla \cdot \hat n\right)^2 + K/2\;
\left(\nabla_{\perp}^2 u \right)^2  \approx K^\prime/2\;
\left(\nabla_{\perp}^2 u \right)^2 $. In the following we drop the
prime and call the new elastic constant $K$.
The
approximated version of $\epsilon_{sym}$ is now given by:
\begin{eqnarray}
  \label{eq:ener_den_sym}
   \epsilon_{sym} & = & \hspace{3ex}
       \frac 12 K\left(\nabla^2_{\perp} u\right)^2
       \nn \\ &&{}
       + \frac 12 B_0 \left[\nabla_z u + (1-n_z) - 
         \frac 12 \left( \nabla_{\perp} u\right)^2 \right]^2 
       \nonumber \\
   &&{}+ \frac 12 B_1 \left(\hat n \times \hat p\right)^2 
\end{eqnarray}
In our model the moduli of the nematic and smectic order parameter
play similar roles, so we will deal with both. 
Since we consider a situation
beyond the phase transition regime, the equilibrium value of the order
parameter is non-zero ($S^{(n,s)}_0$, for both nematic and smectic)
and only its variations $s^{(n,s)}$ can enter the energy density
($S^{(n,s)} = S^{(n,s)}_0 + s^{(n,s)}$). 
\begin{eqnarray}
  \label{eq:ener_den_nord}
   \epsilon_{ord}^{(n)} & = & 
       \frac 12 L_0 \left(s^{(n)}\right)^2 
       + \frac 12 L_{1,ij}^{(n)} 
         \left(\nabla_i s^{(n)}\right) \left(\nabla_j s^{(n)}\right)
       \nn \\ &&{}
       + M_{ijk} \nabla_j n_i \nabla_k s^{(n)} \\[2ex]
  \label{eq:ener_den_sord}
   \epsilon_{ord}^{(s)} & = & 
       \frac 12 L_0 \left(s^{(s)}\right)^2 
       + \frac 12 L_{1,ij}^{(s)} 
         \left(\nabla_i s^{(s)}\right) \left(\nabla_j s^{(s)}\right)
\end{eqnarray}
%
\begin{table}[tbp]
  \begin{center}
    \begin{ruledtabular}
    \begin{tabular}[t]{lcc}
      name &  variable & conjugate \\[1ex]
      \hline
      mass density & $\rho$ & $\mu$ \\[1ex]
      momentum density & $\vec g$ & $\vec v$ \\[1ex]
      nematic director & $\hat n$ & $\vec h$ \\[1ex]
      smectic layer displacement & $u$ & $\Psi$ \\[1ex]
      \begin{minipage}[c]{4cm}
        \raggedright
        variation of the modulus
        of the order parameter
        (either nematic or smectic)
      \end{minipage}
      $\Bigg\}$
      & $s^{(n,s)} $& $\Xi^{(n,s)} $\\
    \end{tabular}  
    \end{ruledtabular}  
    \caption[Variables and their conjugates]
        {Variables and their conjugates, i.e. the corresponding
          thermodynamic forces}
    \label{tab:variables}
  \end{center}
\end{table}

By a similar construction we write down the dissipation function as (see
Tab. \ref{tab:variables} for a list of the thermodynamic variables and
their conjugates) 
\begin{eqnarray}
  \label{eq:dis_coll}
  R & = & R_{cons} + R_{sym} + R_{ord}\\
  \label{eq:dis_det_ani}
  R_{cons} & = & \frac 12 \nu_{ijkl} 
                 \left(\nabla_j v_i\right)  \left(\nabla_l v_k\right) 
                 + R_0\\
  \label{eq:dis_det_sym}
  R_{sym} & = & 
    \frac{1}{2\gamma_1} h_i \delta_{ij}^{\perp} h_j 
        + \lambda_p \Psi^2\\
  \label{eq:dis_det_ord}
  R_{ord} & = & \frac 12 \alpha^{(n)} \; {\Xi^{(n)}}^2 
        + \frac 12 \alpha^{(s)} \; {\Xi^{(s)}}^2
\end{eqnarray}
where $R_0$ summarizes further terms due to conservations laws, which
do not enter our further calculation, and (after \cite{forster71})
\begin{eqnarray}
  \nu_{ijkl} & = & \hspace{3ex}
    \nu_2(\delta_{jl}\delta_{ik} + \delta_{il}\delta_{jk})
    \nonumber \\ && {}
    + 2(\nu_1 + \nu_2 - 2\nu_3)n_i n_j n_k n_l
    \nonumber \\ && {}
    + (\nu_3 - \nu_2)(n_j n_l\delta_{ik} + n_j n_k\delta_{il}
    \nonumber \\ && {}
    \hspace{12ex}      + n_i n_k \delta_{jl} + n_i n_l \delta_{jk})
    \nonumber \\ {} && {}
    + (\nu_4-\nu_2) \delta_{ij} \delta_{kl}
    \nonumber \\ && {}
    + (\nu_5 - \nu_4 + \nu_2)(\delta_{ij}n_k n_l + \delta_{kl}n_i n_j) 
  \label{eq:nu}
\end{eqnarray}
As mentioned in Sec.~\ref{sec:mod} we consider a shear induced
smectic $C$ like situation (but with a small tilt angle, i.e. a weak
biaxiality). We neglect this weak biaxiality in the viscosity tensor
and use it in the uniaxial formulation given above (with the director
$\hat n$ as the preferred direction). This assumption is justified by the
fact that the results presented in this paper do not change
significantly if we use $\hat p$ instead of $\hat n$ in the viscosity
tensor. 

Throughout our calculations, we will not assume any restriction 
on the viscosity constants except the usual requirements due to
thermodynamic stability (see e.g. \cite{pleiner96}). 
Later on we will impose the incompressibility of the fluid by assuming
a constant mass density $\rho$ of the fluid. We emphasize that this
procedure does not require any further assumption about the
material parameters.

The set of basic equations is
completed by the Gibbs-Duhem (the local formulation of the second law
of thermodynamics) and the Gibbs relation (which connects the pressure 
$P$ with the other thermodynamic quantities), which we will use 
in the following form:
\begin{eqnarray}
  \label{eq:gibbs}
  \dd\epsilon &  = & \hspace{3ex}
    \dd\epsilon_0 + \vec v \dd \vec g 
    + \varphi_{ij} \dd \nabla_j n_i + h_i'\dd n_i 
    \nn \\ &&{}
    + \Xi^{\prime (n)} \dd s^{(n)} 
    + \Xi_i^{\prime \prime (n)} \dd \nabla_i s^{(n)}
    \nn \\ &&{}
    + \Xi^{\prime (s)} \dd s^{(s)} 
    + \Xi_i^{\prime \prime (s)} \dd \nabla_i s^{(s)}\\[1ex]
  \label{eq:gibbs_duhem}
  P & = &{} - \epsilon + \mu \rho + T \sigma + \vec v \cdot \vec g 
\end{eqnarray}
The newly defined quantities in Eq.~(\ref{eq:gibbs}) are connected to
the thermodynamic forces (Tab. \ref{tab:variables}) by the following
relations: 
\begin{eqnarray}
  \label{eq:h}
  h_i & = & h_i' - \nabla_j \varphi_{ij} 
        = \frac{\delta \epsilon}{\delta n_i}\\
  \label{eq:psi}
  \Psi & = & - \nabla_i \psi_i 
        = \frac{\delta \epsilon}{\delta u}\\
  \label{eq:xi}
  \Xi^{(n,s)} & = & \Xi^{\prime (n,s)} - \nabla_i \Xi_i^{\prime \prime (n,s)}
        = \frac{\delta \epsilon}{\delta s^{(n,s)}}
\end{eqnarray}

Following the standard procedure within the framework of irreversible
thermodynamics we find the following set of macroscopic hydrodynamic
equations \cite{degroot62,liu79,pleiner96,auernhammer00}. 
\begin{eqnarray}
  \label{eq:u}
  \lefteqn{
  \dt u + v_j \nabla_j u} \;\;\;\;\;\;\; && \nn \\
        & = & v_z - \lambda_p \Psi \\
  \label{eq:n_cart}
  \lefteqn{
  \dt n_i + v_j \nabla_j n_i}\;\;\;\;\;\;\; && \nn \\
        & = & \hspace{3ex}
        \frac 12 \Big[(\lambda-1) \delta_{ij}^{\perp} n_k 
        + (\lambda + 1) \delta_{ik}^{\perp} n_j \Big]\nabla_j v_k
        \nn \\ && {}
        - \frac{1}{\gamma_1} \delta_{ik}^{\perp} h_k \\[1ex]
  \label{eq:kg}
  0 & = & \nabla_i v_i \\
  \label{eq:velo}
  \lefteqn{
  \rho \left(\dt v_i + v_j \nabla_j v_i \right)}\;\;\;\;\;\;\; && \nn \\
        & = & {}
        - \nabla_j \bigg\{\psi_j (\nabla_i u +\delta_{i3}) 
                + \beta_{ij}^{(n,s)}\Xi^{(n,s)} 
                \nn \\ && {} \hspace{6ex}
                - \frac 12 \left[(\lambda-1) \delta_{jk}^{\perp} n_i 
                + (\lambda + 1) \delta_{ik}^{\perp} n_j\right] h_k
                \nn \\ && {} \hspace{6ex}
                + \nu_{ijkl} \nabla_l v_k
                \bigg\} \nn \\ &&{}
        - \nabla_i P\\
  \label{eq:order}
  \lefteqn{
  \dt s^{(n,s)} + v_j \nabla_j s^{(n,s)}}\;\;\;\;\;\;\; && \nn \\
         & = & {}
        - \beta_{ij}^{(n,s)} \nabla_j v_i - \alpha^{(n,s)} \Xi^{(n,s)}
\end{eqnarray}
For the reversible parts of the equations some coupling constants
have been introduced: The flow-alignment tensor 
\begin{equation}
\label{eq:lambda}
\lambda_{ijk} = \frac 12 \left[(\lambda-1) \delta_{ij}^{\perp} n_k +
        (\lambda + 1) \delta_{ik}^{\perp} n_j \right]
\end{equation}
with the flow-alignment
parameter $\lambda$ (and using $\delta_{ij}^\perp = \delta_{ij} -
n_i n_j$) and the coupling between flow and order parameter
\begin{eqnarray}
  \label{eq:beta_n}
  \beta^{(n)}_{ij} & = & \beta_\perp^{(n)} \delta_{ij}^\perp
    + \beta_\parallel^{(n)} n_i n_j \\
  \label{eq:beta_p}
  \beta^{(s)}_{ij} & = & \beta_\perp^{(s)} (\delta_{ij} - p_i p_j)
    + \beta_\parallel^{(s)} p_i p_j. 
\end{eqnarray}
Furthermore there is a reversible coupling between the layer
displacement and the velocity field in equation (\ref{eq:u}). But its
coupling constant has to be unity due to the Gallilei invariance of
the equations. As mentioned above, the use of $u$ is limited to small
layer deformations.

The transverse Kronecker symbols $\delta_{ij}^\perp$ in
Eqs.~(\ref{eq:n_cart},\ref{eq:lambda}) guarantee the 
normalization of $\hat n$. This implies that only two of the
Eqs. (\ref{eq:n_cart}) are independent. 
For the following calculations it turned out to be useful to guarantee
the normalization of the director by introducing two angular
variables $\theta$ and $\phi$ to describe the director. 
\begin{eqnarray}
  \label{eq:dir_sher_1}
  n_x & = & \sin \theta \cos \phi  \\
  \label{eq:dir_sher_2}
  n_y & = & \sin \theta \sin \phi  \\
  \label{eq:dir_sher_3}
  n_z & = & \cos \theta
\end{eqnarray}
Consequently, the Eqs. (\ref{eq:n_cart}) have to be replaced using
angular variables. Denoting the right hand side of
Eqs.~(\ref{eq:n_cart}) with $Y_i$, this can be done the following way.
\begin{eqnarray}
  \label{eq:theta}
  \dt \theta + v_j \nabla_j \theta & = & 
        Y_x \cos \theta  \cos \phi  + Y_y \cos \theta  \sin \phi 
        \nn \\ && {}
        - Y_z \sin \theta  \\
  \label{eq:phi}
  \dt \phi + v_j \nabla_j \phi & = & {}
        - Y_x \frac{\sin \phi }{\sin \theta } 
        + Y_y \frac{\cos \phi }{\sin \theta } 
\end{eqnarray}
In the same way, we guarantee the normalization of $\hat p$ by using
\begin{eqnarray}
  \label{eq:p_norm_1}
  p_x & = & 0\\
  \label{eq:p_norm_2}
  p_y & = & - \nabla_y u\\
  \label{eq:p_norm_3}
  p_z & = & \sqrt{1-p_y^2}
\end{eqnarray}
The different ways of normalizing $\hat n $ and $\hat p$ arise from
the fact, that $\hat p$ is parallel to $\hat e_z$ in zeroth order,
whereas $\hat n$ encloses a finite angle with $\hat e_z$ for any given
shear rate.  

The set of macroscopic hydrodynamic equations we now deal with
(\ref{eq:u}, \ref{eq:kg} -- \ref{eq:order},
\ref{eq:theta}, \ref{eq:phi}) follows 
directly from the initial input in the energy density and the
dissipation function without any further assumptions. 

To solve these
equations we need suitable boundary conditions. In the following we
will assume that the boundaries have no orienting effect on the
director (the homeotropic alignment of the director is only due to the 
layering and the coupling between the layer normal $\hat p$ and the
director $\hat n$). Any variation of the layer displacement must
vanish at the boundaries.
\begin{equation}
\label{eq:bound_u}
u(\pm \frac 12d) = 0
\end{equation}
For the velocity field the situation is a little more complex: 
We assume no-slip boundary conditions, i.e. the velocity of the fluid
and the velocity of the plate are the same at the surface of the
plate. It is convenient to split the velocity field in two parts:
the shear field $\vec v_0$ which satisfies the governing equations and
the no-slip boundary condition and the correction $\vec v_1$ to this
shear field. The boundary condition for $\vec v_1$ now reads:
\begin{equation}
\label{eq:bound_velo_gen}
\vec v_1(\pm \frac 12d) = 0.
\end{equation}
Making use of the following considerations this condition can be
simplified. Due to Eq.~(\ref{eq:u}) the $z$-component of $\vec
v_1$ is suppressed by a factor of $\lambda_p$ (which is typically
extremely small \cite{degennes93,oswald82b}). 
Making use of the results of \cite{oswald82b} we can
assume that $\vec v_1$ depends only on $y$ and $z$ and thus conclude
[with Eq.~(\ref{eq:kg})] that also the $y$-component of $\vec v_1$ is
also suppressed by $\lambda_p$. 
For this reason one can
assume that $v_{1,y}$ and $v_{1,z}$ are negligible and the only relevant
boundary condition for the velocity field is
\begin{equation}
\label{eq:bound_vx}
v_{1,x} = 0.
\end{equation}
The validity of this assumption is nicely illustrated by our
results. Figure \ref{fig:scan-lp-amp} shows that $v_y$ and $v_z$ are
indeed suppressed by $\lambda_p$.

\subsection{Technique of solution}
\label{sec:techn-points}
The aim now is twofold: Finding a spatially homogeneous solution of the
governing equations (for a given shear rate) and investigating the
stability of this solution. In this section we will describe the
general procedure and give the results in Sec. \ref{sec:results}. 

We write the solution as the vector $\vec X = (\theta, \phi,
u, v_{x}, v_{y}, v_{z}, P, s^{(n,s)}) $ consisting of the
angular variables of the director, the layer displacement, the
velocity field, the pressure and the modulus of the (nematic or
smectic) order parameter.
For a spatially homogeneous situation the equations simplify
significantly and the desired 
solution $\vec X_0$ can directly be found (see
Sec.~\ref{sec:hom-state}).  
To determine the region of stability of $\vec X_0$ we perform a
linear stability analysis. I.e. we add a small perturbation $\vec X_1$
to the homogeneous solutions $X_0$:
$\vec X = \vec X_0 + \vec X_1$ (with $\vec X_1
\ll \vec X_0$) and linearize the governing equations in the small
perturbations. In short, the solution of the equation $\text{
  \sffamily L} \; \vec X_1 = \dt \vec X_1$ is analyzed. 
Here {\sffamily L} denotes the operator for the linearized set of the
governing equations. 
The ansatz for the unknown quantities must fulfill the boundary
conditions [see the discussion following Eq.~(\ref{eq:bound_u})]
and follow the symmetry scheme given by Tab.~\ref{tab:sym}.
\begin{table}
  \caption{
    If the symmetry under inversion of $z$ is given for one component of
    $\vec X_1$, the symmetry of all other components follows directly
    from the linearized set of equations. Here we give the $z$-symmetry
    of all components assuming that $u$ is an even function of $z$.
    \label{tab:sym}} 
  \begin{center}
    \begin{ruledtabular}
    \begin{tabular}[t]{cc@{\hspace{3ex}}cc} 
      Quantity & $z$-Symmetry& Quantity & $z$-Symmetry\\
      \hline
      $u$ & even& $v_x$ & even\\
      $\theta$ & odd & $v_y$ & odd\\
      $\phi$ & even& $v_z$ & even\\
      $P$ & odd& $s^{(n,s)}$ & odd\\ 
    \end{tabular}
    \end{ruledtabular}
  \end{center}
\end{table}
Assuming an exponential time dependence and harmonic spatial
dependence of $\vec X_1$
\begin{equation}
  \label{eq:ansatz}
  X_{1,i} \sim \exp[(i\omega + \frac{1}{\tau}) t] \;
    \select{\cos(q y)}{\sin(q y) }\;
    \select{\cos(q_z z)}{\sin(q_z z)}
\end{equation}
fulfills all requirements
(with an oscillation rate $\omega$, a growth rate
$1/\tau$ and a wave vector $\vec q = q \hat e_y + q_z \hat e_z$).
In this ansatz we made use of the results by Oswald and Ben-Abraham
\cite{oswald82b}, who have shown that in standard dilated smectic $A$
under 
shear the first instability  will set in with a wave vector along the
neutral direction of the flow ($\vec q \cdot \hat e_x = 0$). 
After inserting the above ansatz in the linearized set of (partial
differential) equations,
a set of coupled linear equations is obtained to determine
$1/\tau$ and $\omega$. 
From the standard smectic $A$ hydrodynamics it is known, that shear
does not destabilize the layers. Since our extended formulation of
the smectic $A$ hydrodynamics is equivalent to the standard smectic
$A$ hydrodynamics for vanishing external fields (e.g. shear rate),
we assume that the layers are stable for low enough
shear rates, i.e.  $1/ \tau < 0$ for small shear rates.
So $1/ \tau =
0$ marks the set of external parameters (shear rate) and  
material parameters above which $\vec X_1$ grows. 
Typically we hold the material parameters fixed and the only
external parameter is the shear rate. The
solvability condition of the corresponding set of linear equations
gives a relation between the shear rate [and tilt angle $\theta_0$,
which is directly connected to the shear rate, see
Eq.~(\ref{eq:theta_0}) below], $\omega$ and the wave vector $q$. For
every given $q$ a specific shear rate (and tilt angle $\theta_0$) can
be determined which separates the stable region (below) from the
unstable region (above). This defines the so called
curve of marginal stability (or neutral curve) $\theta_0(q)$. If, for
any given set of external 
parameters, the tilt angle $\theta_0$ lies above the curve of marginal
stability for at least one value of $q$, the spatial homogeneous state
is unstable and undulations grow.
The smallest shear rate (tilt angle) for which undulations can grow is
called the critical shear rate (tilt angle). Technically speaking, we solve
$\text{\sffamily L} \; \vec X = i \omega \vec X$ --- in many cases we
can set 
$\omega = 0$, see below. We point out that this linear analysis
is only valid at the point where the first instability sets
in. Without further investigations no prediction of the spatial
structure of the developing instability can be made. 
Also the nature of the 
bifurcation (backward or forward) must be determined by further
investigations. 

For practical reasons we used dimensionless units in our numerical
calculations. The invariance of the governing equations under
rescaling time, length and mass allows us to choose three parameters
in these equations to be equal to unity. We will set 
\begin{equation}
  \label{eq:dim-less}
  B_1 = 1\mbox{, } \gamma_1 = 1\mbox{, and } \frac d \pi = q_z = 1
\end{equation}
and measure all other quantities in the units 
defined by this choice. Nevertheless we will keep these quantities 
explicitly in our analytical work.

To extract concrete predictions for experimental parameters from our
calculations is a non-trivial task, because neither the energetic
constant $B_1$ nor the rotational viscosity $\gamma_1$ are used for
the hydrodynamic description of the smectic $A$ phase (but play an
important role in our model).  Therefore,
we here rely on measurements in the
vicinity of the nematic-smectic $A$ phase transition.
Measurements on low molecular weight liquid
crystals made by Litster \cite{litster79} in the
vicinity of the nematic-smectic $A$ transition indicate that $B_1$ is
approximately one order of magnitude less than $B_0$. 
As for $\gamma_1$ we could not find any measurements which would
allow an estimate of its value in the smectic $A$ phase. In the nematic
phase $\gamma_1$ increases drastically towards the nematic-smectic
$A$ transition (see e.g. \cite{graf92}). Numerical simulations on a
molecular scale are also a promising approach to determine these
constants \cite{soddemann02}. 

Due to the complexity of the full set of governing equations, we will
start our analysis with a minimal set of variables ($\theta$, $\phi$
and $u$) and suppress the coupling to the other variables (see
Sec.~\ref{sec:min-vari}).  
Step by
step the other variables will be taken into account. The general
picture of the instability will turn out to be already present in the 
minimal model, but many interesting details will be added throughout
the next sections.
In comparison to our earlier work
\cite{auernhammer00} we now use the way of normalizing $\hat n$ 
and $\hat p$ derived above. 
This will lead to some small differences in the results
but leaves the general picture unchanged. First we assume a stationary
instability (i.e. we let $\omega=0$); later on we discuss the
possibility for an oscillatory instability and have a look at some
special features of the system (Secs.~\ref{sec:oscill-inst}
and \ref{sec:anisotric-viscosity}).

\section{Results and discussion}
\label{sec:results}
\subsection{Spatially homogeneous state}
\label{sec:hom-state}
Looking for a spatially homogeneous solution, the governing equations
simplify significantly. A linear shear profile 
\begin{equation}
  \label{eq:v_0}
  \vec v_0 = \dot \gamma z \hat e_x
\end{equation}
is a solution to (\ref{eq:velo}) and $u$ stays unchanged in this
regime. The only variables which have a zeroth order correction for
all shear rates are the tilt angle $\theta$ and the modulus of the
nematic order parameters $s^{(n)}$:
\begin{eqnarray}
  \label{eq:theta_0}
  \lefteqn{
  \left(\frac{\lambda +1}{2} -\lambda \sin^2(\theta_0) \right) \dot
        \gamma} \;\;\;\;\;\;\;\;\;\;\;\;\;\;\;\;\;&&\nn\\
        & = & \frac{B_1}{\gamma_1} \sin(\theta_0)\cos(\theta_0)
        \nn \\ &&
        + \frac{B_0}{\gamma_1} \sin(\theta_0)(1-\cos(\theta_0)) \;\;
        \\[1ex] 
  \label{eq:s_0_n}
  \alpha^{(n)} L_0 s_0^{(n)}
        & = & (\beta^{(n)}_\parallel - \beta^{(n)}_\perp)
              \sin(\theta_0) \cos(\theta_0) \dot \gamma
\end{eqnarray} 
Equation (\ref{eq:s_0_n}) shows that
nematic degrees of freedom couple to simple shear, but not the
smectic degrees of freedom; the modulus of the nematic order
parameter has a non-vanishing spatially homogeneous correction [see
Eq.~(\ref{eq:s_0_n})], whereas
the smectic order parameter stays unchanged. 
The reason for this difference 
lies in the fact that $\beta^{(n)}_{ij}$ and $\beta^{(s)}_{ij}$
include $\hat n$ and $\hat p$, respectively, which coupled differently
to the flow field [see Eq.~(\ref{eq:beta_n}, \ref{eq:beta_p})].  
Eq.~(\ref{eq:theta_0}) gives a
well defined relation between the shear rate $\dot \gamma$ and
the director tilt angle $\theta_0$, which we will use to eliminate
$\dot \gamma$  from our further  
calculations. To lowest order $\theta_0$ depends linearly on $\dot
\gamma$: 
\begin{equation}
  \label{eq:theta_0_lin}
  \theta_0 = \dot \gamma \frac{\gamma_1}{B_1} \; \frac{\lambda +1}{2}
        + O(\theta^3_0)
\end{equation}
We are not aware of any experimental data, which would allow a direct
comparison with these results. We stress, however, that  molecular
dynamics simulations by Soddemann et al. \cite{soddemann02} are in
very good agreement with Eqs.~(\ref{eq:theta_0}) and
(\ref{eq:theta_0_lin}). 

In contrast to the director tilt the lowest
order correction to the nematic order parameter is quadratic in the
shear rate (tilt angle).
\begin{equation}
\label{eq:s_0_n_lin}
s^{(n)}_0 = \frac{2}{\lambda+1} \; \frac{B_1}{\gamma_1} \;
        \frac{\beta^{(n)}_\parallel - \beta^{(n)}_\perp} {\alpha^{(n)}
          L_0}  \theta^2_0 + O(\theta^4_0)
\end{equation}
In the following we consider perturbations around the spatially
homogeneous state given above.  

\subsection{Stationary instability}
\label{sec:stat-inst}
\subsubsection{Minimal set of variables}
\label{sec:min-vari}
Let us first consider the effect of our modifications regarding the 
normalization of $\hat n$ and $\hat p$ in comparison to our earlier
results \cite{auernhammer00}. 
For this purpose we consider only a minimal set of variables: 
the director (characterized by the two angles $\theta$ and $\phi$) and 
the layer displacement $u$. We neglect all couplings of these
variables to other quantities describing the system, namely 
the velocity field and the moduli of the nematic and smectic order
parameters. Within these approximations the equations to solve are: 
\begin{eqnarray}
0 & = & \hspace{3ex}A_{\theta}
  \begin{array}[t]{l}
    \Big\{
    2 \dot \gamma \lambda \sin(\theta_0) \cos(\theta_0)\\\hspace{2ex}
    + \displaystyle
	\frac{B_0}{\gamma_1} \left[\sin^2(\theta_0) - \cos^2(\theta_0) +
    \cos(\theta_0)\right]
    \\[3ex]\hspace{2ex}
    - \displaystyle
	\frac{B_1}{\gamma_1} \left[\sin^2(\theta_0) -
	\cos^2(\theta_0)\right] \Big\}  
  \end{array} \nn \\
& & {}- A_u\;\; \frac{B_0}{\gamma_1}\sin(\theta_0)q_z
  \label{eq:min-var-t}
  \\[2ex]
0 & = & \hspace{3ex}
  A_{\phi} \;\;
  \frac 12 \dot \gamma (\lambda+1)
  - A_u \;\;\frac{B_1}{\gamma_1} q
  \label{eq:min-var-f}
  \\[2ex]
0 & = &\hspace{3ex}
  A_{\theta} \;\;
  \lambda_p B_0 \sin(\theta_0) q_z 
  \nn\\
  &&{}
  + A_{\phi} \;\;\lambda_p B_1 q\sin(\theta_0) \cos(\theta_0)
  \nn\\
  &&{}  
  -A_u \;\; \lambda_p
    \Big[
    -B_0 {q}^{2}(1-\cos(\theta_0))
    \nn \\ &&{}\hspace{9ex}
    +B_1{q}^{2}\cos^{2}(\theta_0)
    +K{q}^{4}
    +B_0 q_z^2
    \Big]
  \label{eq:min-var-u}
\end{eqnarray}
Here we inserted an ansatz of the type (\ref{eq:ansatz}) and denoted
the linear amplitudes of $\theta$, $\phi$, and $u$ by
$A_{\theta}$, $A_{\phi}$, and $A_u$, respectively.
One can solve these equations either by expanding them in a power
series of $\theta_0$ (expecting to get a closed result for the
critical values) or numerically. It turns out, 
that one has to take into account terms (at least) up to order
${\theta_0}^5$ in Eqs.~(\ref{eq:min-var-t} -- \ref{eq:min-var-u}) to
get  physically meaningful (but rather long and complicated)
analytical results.  For this reason the closed expressions have
no advantage over the purely numerical solutions and we do not
give the analytical approximations explicitly. A comparison with the
results of Ref.~\cite{auernhammer00} will be given in
App.~\ref{sec:analyt-appr-minim}.
We will present and discuss our findings using the minimal set of
variables in Sec.~\ref{sec:coupl-velo} in direct comparison to the
results of the full set of equations.

\subsubsection{Coupling to the velocity field}
\label{sec:coupl-velo}
In the previous section we have shown that already a minimal set of
variables supports our picture of the physical mechanism. But
neglecting the coupling between velocity field and nematic director and 
vice versa is a rather crude approximations since it is well known,
that this coupling plays an important roll in nematohydrodynamics
\cite{degennes93,pleiner96}. So the natural next step is to include
this coupling and to perform a linear
stability analysis of Eqs.~(\ref{eq:u} -- \ref{eq:velo},
\ref{eq:theta}, \ref{eq:phi}). In this case the  
standard procedures leads to a system of seven coupled linear
differential equations. Following the discussion after
Eq.~(\ref{eq:bound_u}) these equations can be solved by an
ansatz of the type given in Eq.~(\ref{eq:ansatz}). This reduces the 
system of equations to seven coupled linear equations which are easily
solved using standard numerical tools (such as singular value 
decomposition and inverse iteration to find the eigenvectors). Due to
the complexity of the equations we used Maple to determine the final
set of linear equations. The key ingredients of this Maple script are
given in App.~\ref{sec:details-maple}. 

\begin{figure}
\caption[]{\label{fig:nk1-normalfall}
  A typical picture for the comparison of the neutral curves using  the
  minimal set of variables (\includegraphics*{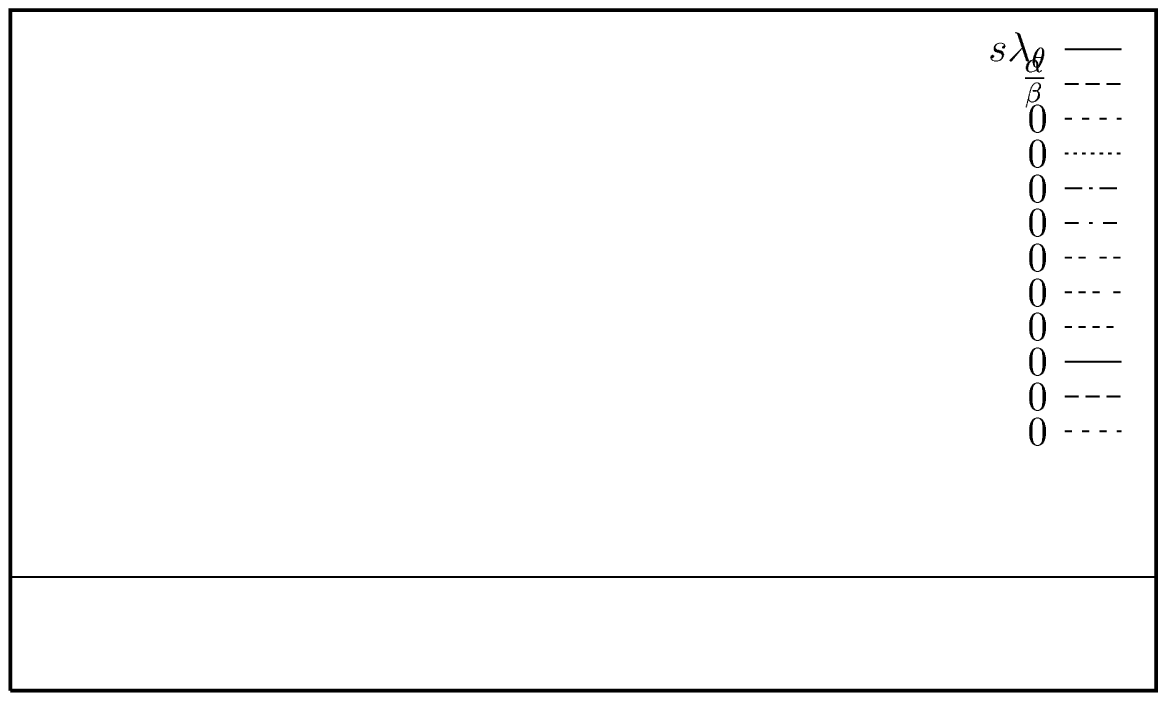})
  and including the velocities field (\includegraphics*{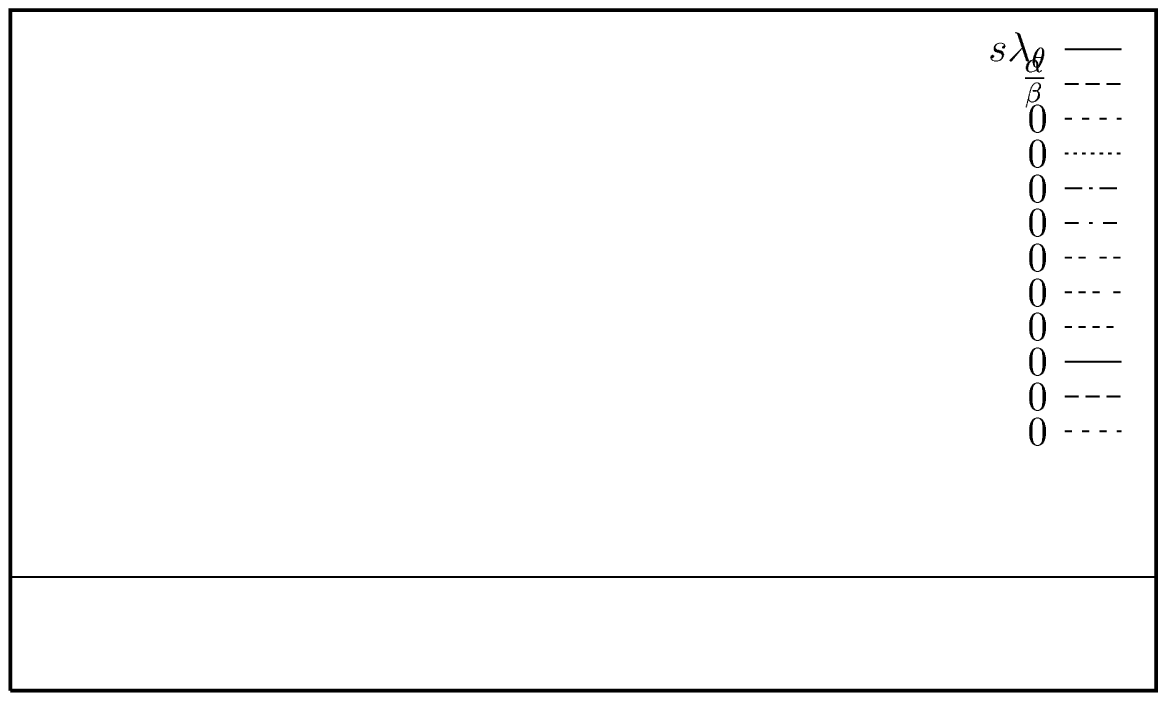}). The
  overall behavior does not change but the critical values are altered 
  due to the coupling with the velocity field. For this plot we used
  (in the dimensionless units discussed in
  Sec.~\ref{sec:techn-points}) $B_0 = 10$, $K = 
  10^{-6}$, $\lambda = 1.1$, $\nu_1= \nu_2= \nu_3= \nu_4= \nu_5 =
  0.1$ and $\lambda_p = 10^{-6}$. 
  The inset shows the linear amplitudes $A_i$ (where $i$ stands for
  $\theta$, $\phi$, etc) at onset. Since the logarithm
  of the amplitudes is shown, amplitudes with different sign are shown
  with a different line
  style. Using the minimal set (left bars) all
  amplitudes have the same sign
  (\includegraphics*{line1.ps}). Including the velocity field 
  (right bars) some
  amplitudes are positive (\includegraphics*{line2.ps}), others
  negative (\includegraphics*{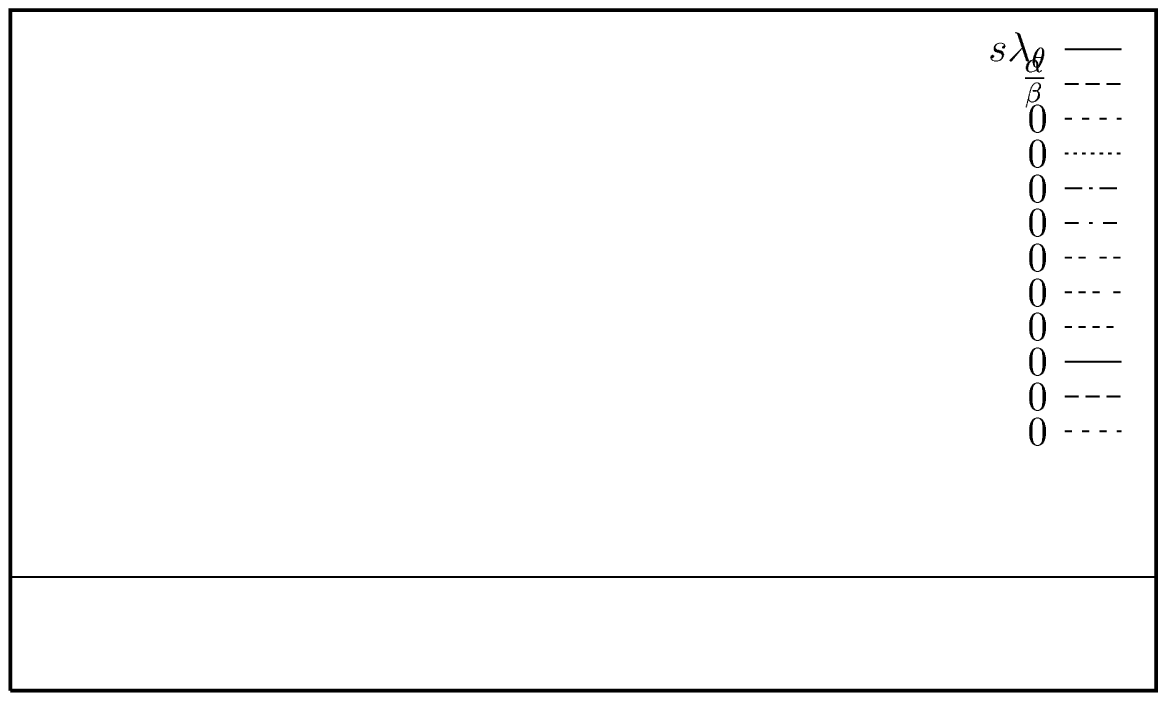}).
  Note that we use in this and all following plots the dimensionless
  units defined by Eq.~(\ref{eq:dim-less}). 
}
\center\includegraphics[width=\columnwidth]{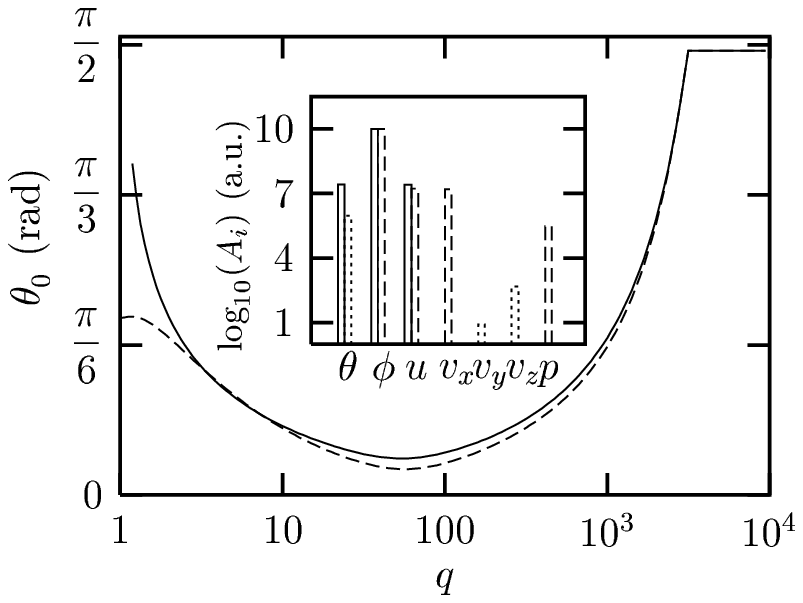}
\end{figure}

Figure \ref{fig:nk1-normalfall} gives a comparison of typical neutral
curves for the minimal model and calculations including the velocity
field. The overall shape of the neutral curve
is not changed due to the coupling to the velocity field but a shift
of the critical values (especially in the critical tilt angle)
is already visible. The inset shows the
relative amplitudes of the linear solutions at onset on a logarithmic
scale. For $\theta$, $\phi$ and $u$ the left bars correspond to the
minimal model and the right bars to the extended version.
Note that amplitudes with a different sign are shown with a different
line style in the histograms (see figure caption for details).

\begin{figure}
  \caption[]{\label{fig:scan-b}
    A significant difference between the various approaches is only
    visible for $B_0 \lesssim 100$. At higher values of $B_0$ the
    number of free variables plays no noticeable role and the critical
    values follow a master curve. The solid lines
    (\includegraphics*{line1.ps}) show 
    results including the velocity field, the dashed lines
    (\includegraphics*{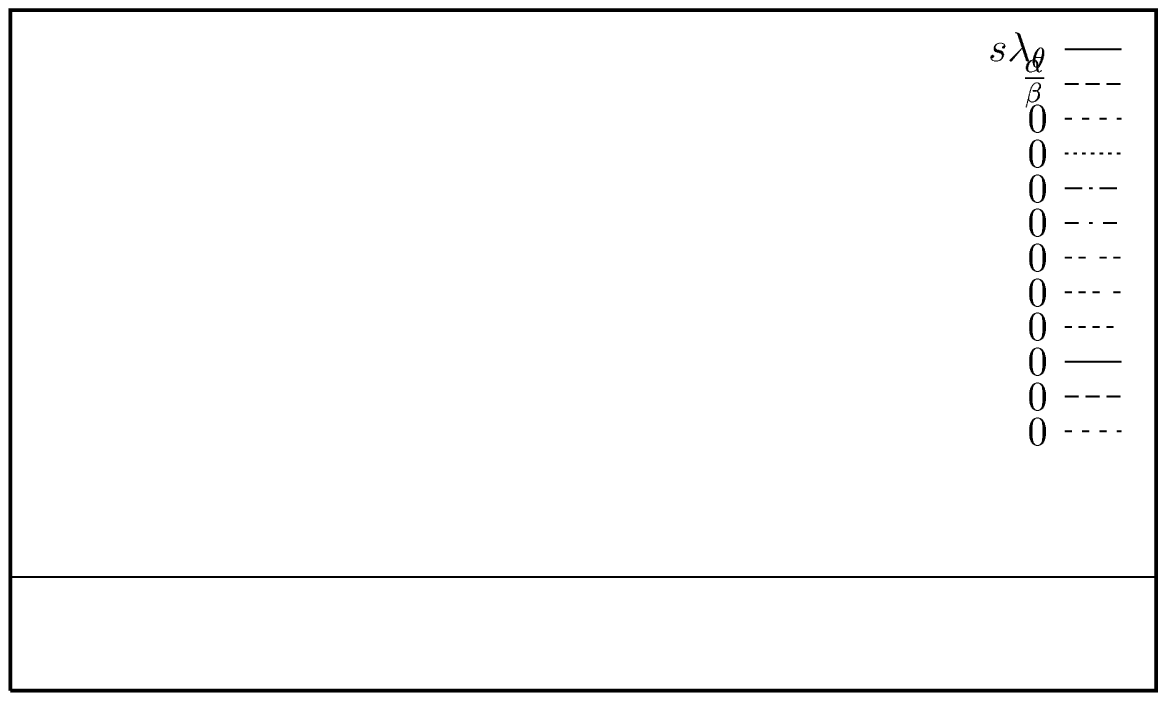}) correspond
    to the minimal set of variables. At low $B_0$ in the upper curves
    we used $\lambda = 2$.
}
\center\includegraphics[width=\columnwidth]{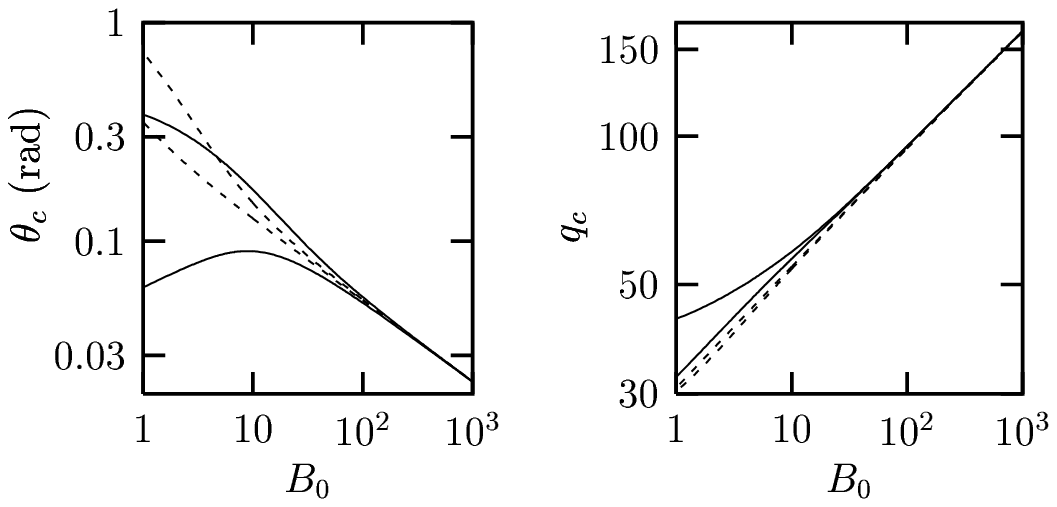}
\end{figure}

\begin{figure}
  \caption[]{\label{fig:scan-k}
    Plotting the critical values as a function of the bending modulus
    $K$ shows a convergence of the curves, which is nevertheless 
    not as pronounced as in the case of Fig.~\ref{fig:scan-b}. The
    influence of $\lambda$ on the critical tilt angle is 
    significant ($\lambda = 2$ in the upper curves and $\lambda = 1.1$ in
    the lower ones). Again the solid lines (\includegraphics*{line1.ps}) show
    results including the velocity field and the dashed lines
    (\includegraphics*{line3.ps}) correspond
    to the minimal set of variables. 
}
\center\includegraphics[width=\columnwidth]{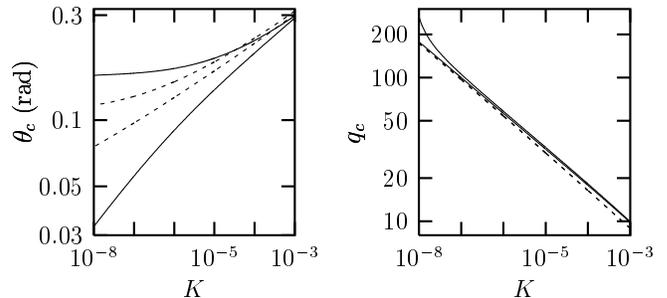}
\end{figure}

\begin{figure}
  \caption{\label{fig:scan-lp-amp}
    In all our calculations $v_{1,x}$ is the dominating component of
    $\vec v_1$. This graph demonstrates that the other components are
    suppressed by $\lambda_p$ (making them almost negligible).
    }
  \center\includegraphics[width=5cm]{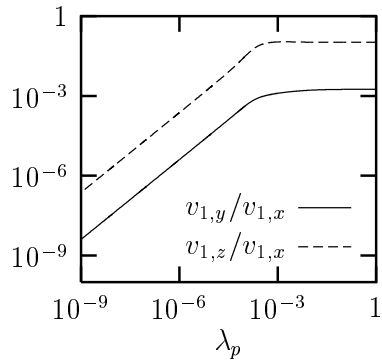}
\end{figure}

\begin{figure}
  \caption[]{\label{fig:scan-nu}
    Only the viscosities $\nu_2$ and $\nu_3$ can influence the
    critical parameters significantly. The upper row depicts the
    dependence on a isotropic variation of the viscosity. In the
    middle and lower row we present the variation with $\nu_2$ and
    $\nu_3$ setting the other viscosities to $\nu_i = 0.1$. Here the
    thick solid lines (\includegraphics*{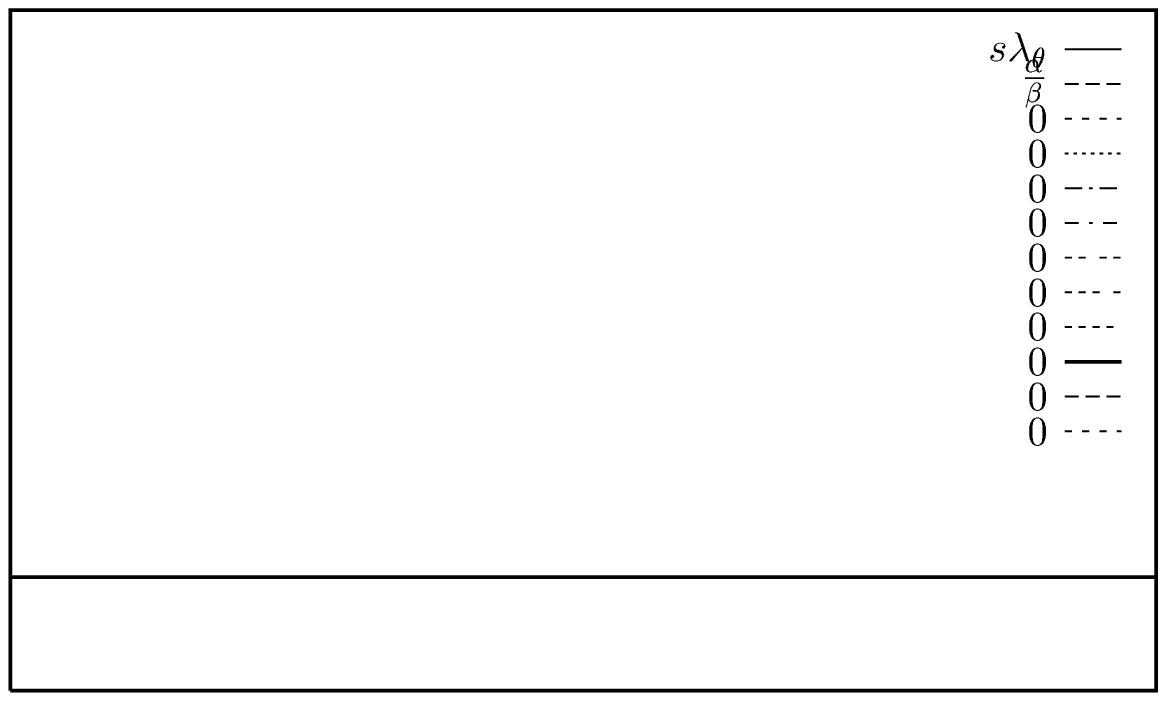}) represent the
    minimal set of variables. For the full set of variables we have
    chosen four different values of $\lambda$: the solid curves  
    (\includegraphics*{line1.ps}) with $\lambda = 0.7$, the dashed
    curves (\includegraphics*{line2.ps}) with $\lambda = 1.1$, 
    the dotted curves (\includegraphics*{line4.ps}) with $\lambda = 2$
    and the dot-dashed curves (\includegraphics*{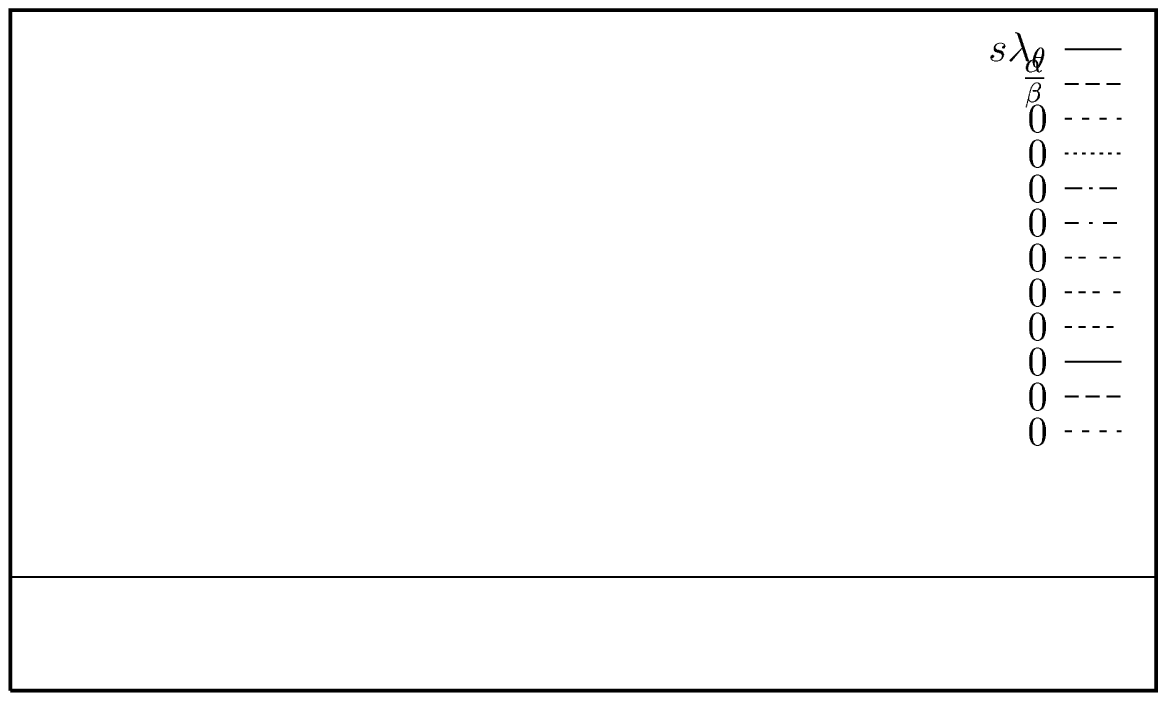}) with
    $\lambda = 3.5$. Note the similarities between the curves for
    small (\includegraphics*{line1.ps}) and large $\lambda$
    (\includegraphics*{line5.ps}) in the upper and middle row: In
    these regimes $\nu_2$ is the dominating viscosity.
    }
  \center\includegraphics[width=\columnwidth]{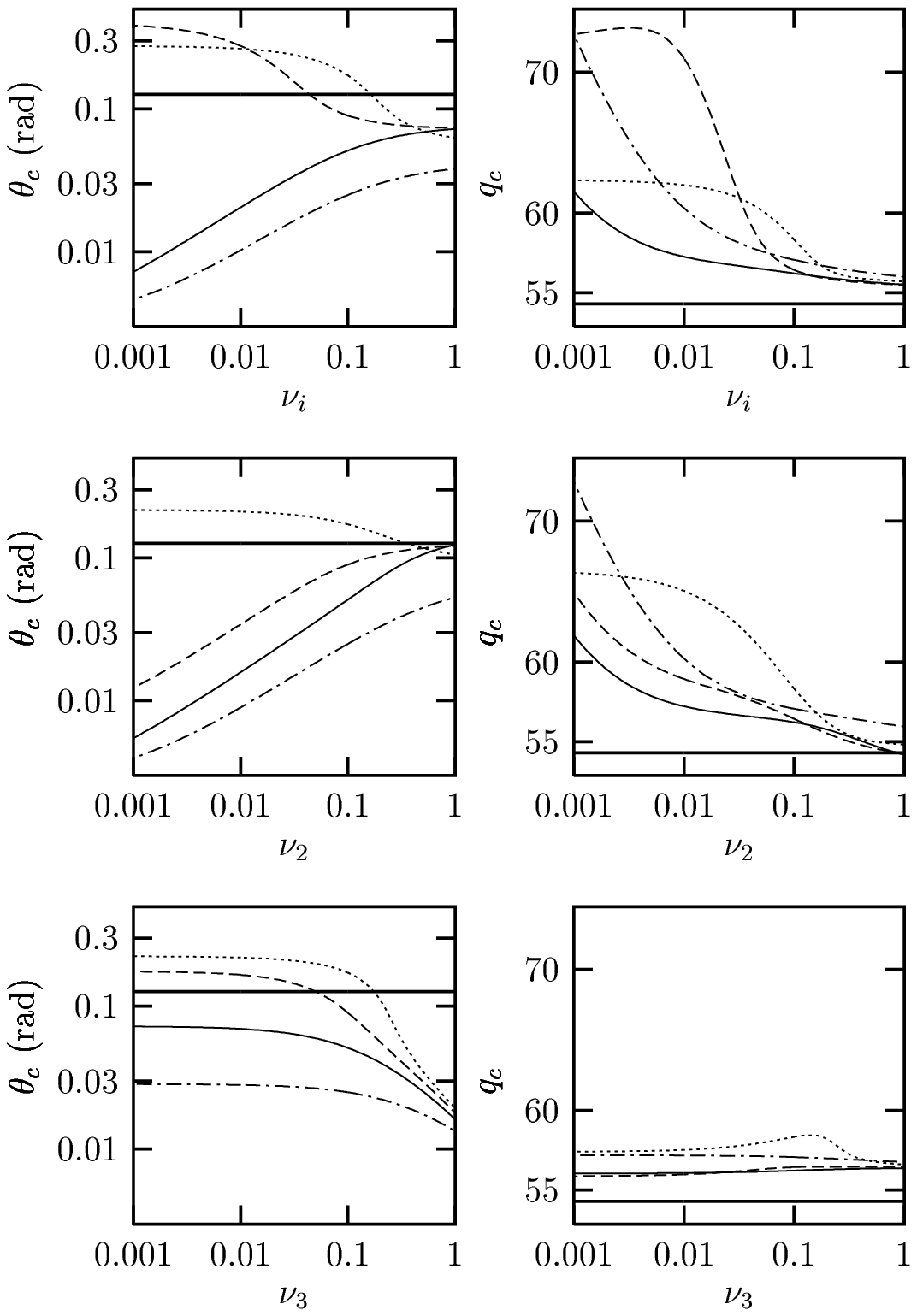}
\end{figure}

\begin{figure}
  \caption[]{\label{fig:scan-lam}
    Plotting the critical values as functions of the flow alignment
    parameter $\lambda$ reveals an interesting structure for $1
    \lesssim \lambda \lesssim 3$. 
    In the upper row we plot this dependence for a set 
    of (isotropic) viscosities ranging from $\nu_i= 1$ (thick solid line,
    \includegraphics*{line10.ps}) down to $\nu_i = 10^{-3}$ (thick dashed
    line, \includegraphics*{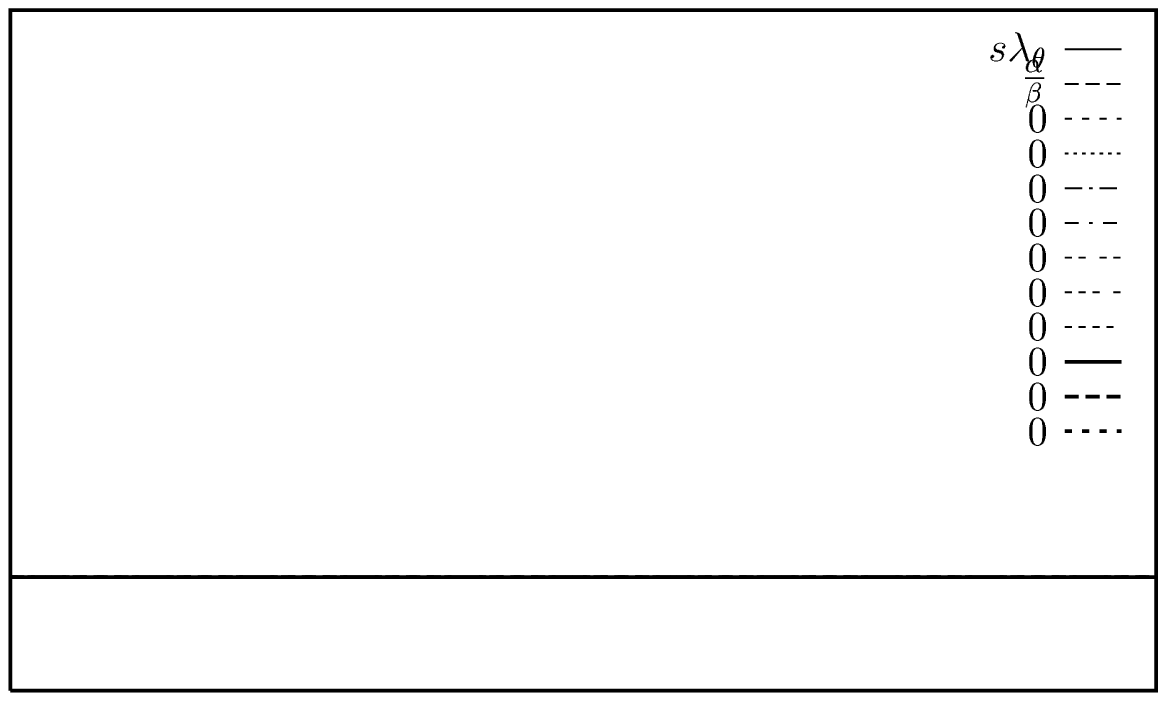}).
    The dotted line
    (\includegraphics*{line4.ps}) reveals that this dependence on
    $\lambda$ is absent in the minimal model. 
    The lower row illustrates the behavior for varying layer
    compressibility $B_0$ with $B_0 \approx 3$ for the thick solid curve
    (\includegraphics*{line10.ps}) and $B_0 = 100$ for the thick dashed
    curve (\includegraphics*{line11.ps}). In all plots the thin solid
    lines (\includegraphics*{line1.ps}) give the behavior for some
    intermediate values.
    For an interpretation of this behavior see the text. 
    }
  \center\includegraphics[width=\columnwidth]{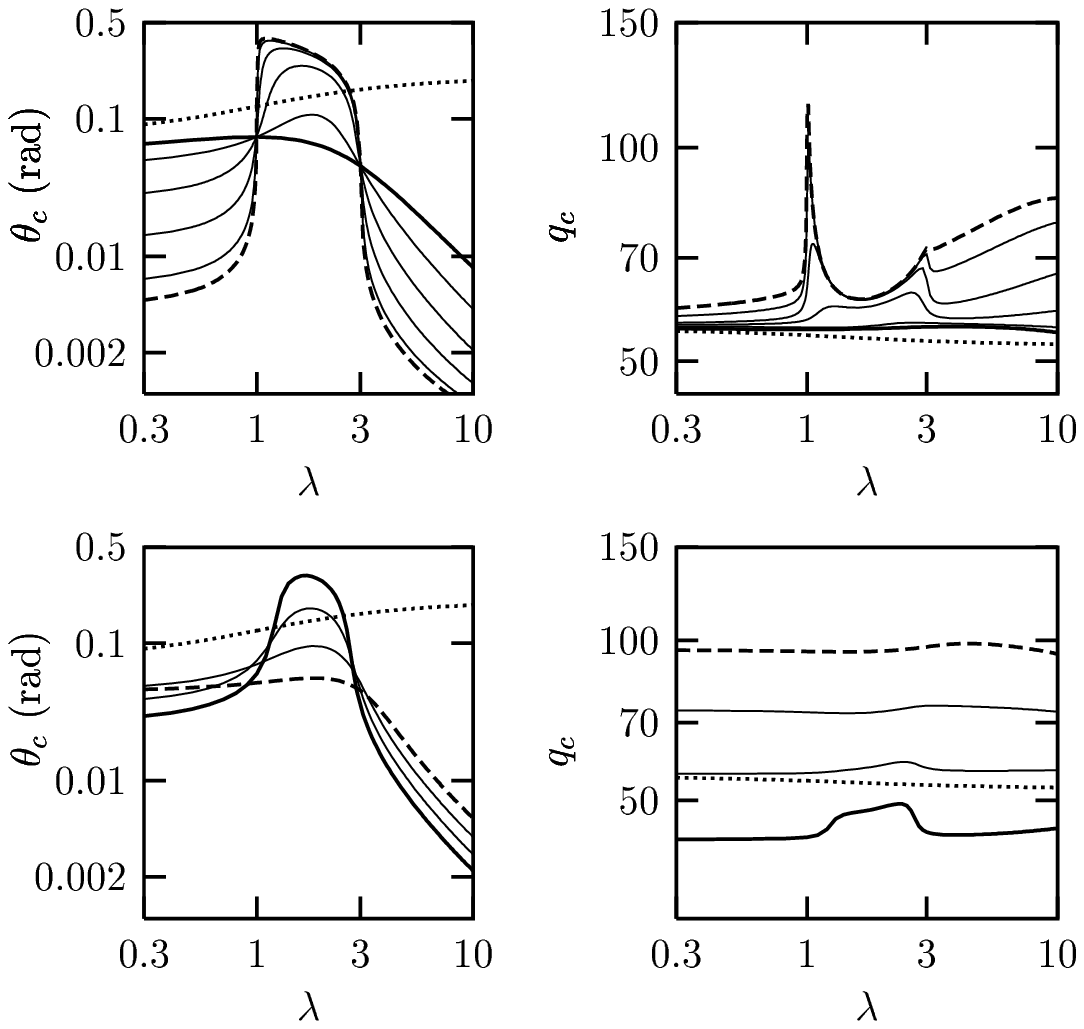}
\end{figure}

Let us have a closer look
at the differences between the minimal and the extended set of
equations and follow these differences along some paths in the
parameter space. As mentioned in Sec.~\ref{sec:techn-points}, we can
omit some of the physical parameters by using dimensionless
parameters. In Figs.~\ref{fig:scan-b} -- \ref{fig:scan-lam} we show
the dependence of the critical values of the tilt angle and wave
vector on the dimensionless
parameters [as defined in Eq.~(\ref{eq:dim-less})].
For all these figures we used the same basic set of parameters: 
$B_0 = 10$, $K = 10^{-6}$, $\lambda = 1.1$, $\nu_1= \nu_2= \nu_3=
\nu_4= \nu_5 = 0.1$  and $\lambda_p = 10^{-6}$. These values are
estimates for a typical thermotropic LMW liquid crystal, where we made
use of the results of \cite{litster79,graf92} (as far as $B_1$ and
$\gamma_1$ are concerned, see also the last paragraph in
Sec.~\ref{sec:min-vari}). 
For flow alignment parameters in the range $1 \lesssim \lambda
\lesssim 3$ the critical values vary strongly with $\lambda$
(see Fig.~\ref{fig:scan-lam}). Therefore we discuss in addition
the situation for $\lambda = 2$ to indicate the range of possible
values. 

Considering the critical values as a function of the compression
modulus $B_0$ results in a rather simple situation
(Fig.~\ref{fig:scan-b}): For small values 
of $B_0$ a significant influence of the coupling between the director
and velocity field is apparent, which also shows a strong dependence
on $\lambda$. For large $B_0$ all these differences vanish and only
one single curve is obtained. At this point a comparison to dilated
smectic $A$ is instructive. It is well known \cite{delaye73,clark73}
that in dilated smectic the critical wave vector and the critical
dilatation show a power law behavior as a function of $B_0$ with
exponents $1/4$ and $-1/2$, respectively. In the limit of large $B_0$
we found the same exponents already in our earlier analysis
\cite{auernhammer00}. If we fit power 
laws to our results for $B_0 > 10^2$ we find the exponents equal to
$\approx 0.235$ and $\approx -0.37$ for $q_c$ and $\theta_c$,
respectively  (note that the dilatation in our model is $\approx \frac
12 \theta_c^2$). 
So both approaches (the minimal set of variables and the calculations 
including the velocity field) show, despite all similarities to the
standard model of smectic $A$ and to our earlier analysis, differences
in the details of the instability. 

A similar, but less pronounced, situation is apparent, when plotting
the critical values as a function of the bending modulus (see
Fig.~\ref{fig:scan-k}). The curves
tend to converge for large $K$, but there remains a difference between 
the minimal set of variables and the calculations including the the
velocity field. Fitting the $K$-dependence with power laws (here for
$K > 10^{-4}$) only the critical wave number exhibits an exponent
close to the values expected from dilated smectic $A$ ($\approx
-0.26$ vs. $- \frac 14$). This illustrates the fact that shearing a
lamellar system is similar to dilating it but not equivalent. 

In contrast to the cases discussed above, the permeation constant
$\lambda_p$ has no strong influence on the critical values. For
dimensionless values $\lambda_p < 10^{-6}$ the critical values do not
change at all with $\lambda_p$. For large values variations within a
factor of two are possible. The
permeation constant is known to be very small. In our dimensionless
units we expect it be of the order of $< 10^{-9}$ for LMW
thermotropic liquid crystals and neglect its influence on the critical
values for this reason.
In Sec.~\ref{sec:imp} we have emphasized that the $y$- and
$z$-components of the velocity field  are suppressed via
$\lambda_p$. These qualitative arguments are clearly confirmed by our
numerical results: In all our calculations $v_{1,x}$ is the dominating
component of $\vec v_1$ and the ratio $v_{1,y}/v_{1,x}$ is of the
order of $\lambda_p$ over the whole range of physical relevant values
of $\lambda_p$ (see Fig.~\ref{fig:scan-lp-amp}).
This fact nicely supports our argument
that we can neglect the boundary condition for $v_{1,y}$, because
$v_{1,y}$ vanishes anyway.

Out of the five viscosities only two ($\nu_2$ and $\nu_3$) show a
significant influence on the critical values. In
Fig.~\ref{fig:scan-nu} we present the dependence of $\theta_c$ and
$q_c$ on an assumed isotropic viscosity (upper row) and on 
these two viscosity coefficients (middle and lower row). Since the
flow alignment parameter $\lambda$ has remarkable influence on these
curves we have chosen four different values of $\lambda$ in this
figure, namely $\lambda = 0.7$, $\lambda = 1.1$, $\lambda = 2$ and
$\lambda = 3.5$. 
The curves for $\lambda \lesssim 1$ and $\lambda \gtrsim 3$ for an
isotropic viscosity tensor
are very similar to the corresponding curves where only $\nu_2$ is
varied. 
In this parameter range the coefficient
$\nu_2$ dominates the behavior. Note that the influence of $\nu_3$ 
on the critical values is already much smaller than that of $\nu_2$.
We left out the equivalent graphs for the other viscosity
coefficients, because they have almost no effect on the critical
values. For further comments on the influence of an anisotropic
viscosity tensor see also Sect.~\ref{sec:anisotric-viscosity}.

All the parameters we have discussed up to now caused variations in
the critical values that did not select specific values of the
considered parameter. In this aspect the situation is completely
different in the case of the flow alignment parameter $\lambda$. As
shown in Fig.~\ref{fig:scan-lam} there is a clear change in behavior
for $\lambda \approx 1$ and $\lambda \approx 3$. The critical tilt
angle is increased for values of $\lambda$ in this interval and the
critical wave vector tends to rise only at the boundaries of the
interval and is reduced in between. 
Fig.~\ref{fig:scan-lam} illustrates how this structure depends on the
viscosities (assuming all five viscosities to be equal) and on the
elastic constants of the layers.
In the first row we follow this behavior for viscosities
varying from $\nu_i = 1$ down to $\nu_i = 10^{-3}$. Clearly, the
influence of $\lambda$ is more pronounced the lower the viscosities
are. 
Both elastic constants of the layers, the compressibility
$B_0$ and the bending modulus $K$ (in our dimensionless units $B_1 =
1$), have in general a similar influence on the shape of the graphs:
the smaller the elastic constants are, the more pronounced the
structure becomes. For this reason we just give the plot for $B_0$
(second row in Fig.~\ref{fig:scan-lam}) and omit the plot for $K$. 

These dependencies on the system parameters give some important hints
for an interpretation of Fig.~\ref{fig:scan-lam}. The currents and
quasi-currents for the velocity field and the director consist of two
parts [see Eqs.~(\ref{eq:n_cart}) and (\ref{eq:velo})]: a diagonal one
(coupling e.g. the components of $\vec v$ among 
each other) and an off-diagonal one (coupling the director to
the velocity field). The former ones are proportional to the elastic
constants or to the viscosity tensor whereas the latter one is a
function of the flow alignment parameter. So reducing either the
elastic constants or the viscosities increases the portion of the
cross-coupling terms in theses equations. I.e. the observed tendencies
are exactly what one would expect. The next step 
in the interpretation of the shape
of the curves is to have a closer look at the structure of the
cross-coupling term. The flow alignment tensor $\lambda_{ijk} = 
\frac 12 \left[(\lambda-1) \delta_{ij}^{\perp} n_k +
        (\lambda + 1) \delta_{ik}^{\perp} n_j \right]$ obviously
changes its behavior for $\lambda = 1$: The first part changes its
sign. Note that we are in a region of the parameter space, where
$\lambda_{ijk}$ is a dominating term (since either the viscosities or
the elastic constants are small). Additionally, $\delta_{ij}^\perp
n_k$  contains up to the third power of one
director component. For this reason we expect that --- in the
linearized set of equations --- some coupling terms 
change their sign for $\lambda = 1$ others for $\lambda = 3$.
E.g. the $\phi$-component of the director is coupled to the $x$-
and $z$ component of the velocity field by the terms $(\lambda
  -1)/2\; \partial_y v_x$ and $(\lambda -1)/2\;
\cot(\theta_0)\partial_y v_z$. Similarly the
reversible part of the coupling of $v_y$ to $\phi$ vanishes for
$\lambda = 3$. The monitored structure in the plots cannot be
attributed 
to one single cross-coupling term, but the given examples demonstrate
that something should happen in this parameter range.

\subsubsection{Including the order parameters}
\label{sec:effects-order}
\begin{figure}
  \caption{\label{fig:scan-s0}
    Evaluating Eqs.~(\ref{eq:s_0_n_lin}) and
    (\ref{eq:limit-alpha}) at onset gives an important restriction on
    the range of possible parameter values (here the cases of
    $\alpha^{(n)}$ and $\beta^{(n)}_\parallel - \beta^{(n)}_\perp$). 
    Note that the critical $\theta_0$ is a function of the material
    parameters.
    }
  \center\includegraphics[width=\columnwidth]{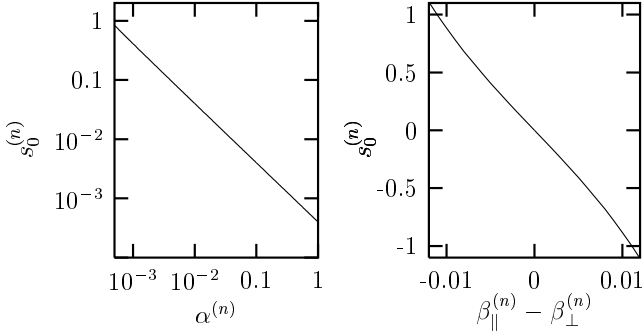}
\end{figure}

%
%
In the preceeding paragraphs we investigated undulations assuming a
constant modulus of the order parameter $S^{(n,s)} = S^{(n,s)}_0 +
s^{(n,s)}_0$. In general one would expect that the undulations in the  
other observable quantities should couple to some extent to the order
parameter. 
In the formulation of the free energy (see
Sec.~\ref{sec:imp}) we have assumed that
$S^{(n,s)}$ varies only slightly around $S^{(n,s)}_0$ and thus only 
the lowest order terms in $s^{(n,s)}$ contribute to the free
energy. For the spatially homogeneous state we had [see
Eqs.~(\ref{eq:s_0_n}, \ref{eq:s_0_n_lin})] a correction to the nematic
$S^{(n)}$ proportional to the square of the shear rate ($\theta_0 \sim
\dot\gamma$ for low $\dot \gamma$):
\begin{displaymath}
  s^{(n)}_0 = {}-\frac{2}{\lambda+1} \; \frac{B_1}{\gamma_1} \;
              \frac{\beta^{(n)}_\parallel - \beta^{(n)}_\perp}
              {\alpha^{(n)} L_0} \theta^2_0 + O(\theta^4_0) 
\end{displaymath}
As a consequence $s^{(n)}_0$ must be small compared to
$S^{(n,s)}_0$ (which is by construction limited to the range $0 \le
S^{(n,s)}_0 \le 1$). Thus a reasonable restriction is 
\begin{equation}
  \label{eq:limit-alpha}
  |s_0^{(n)}| \lesssim 0.5.
\end{equation}
As shown in Fig. \ref{fig:scan-s0}, evaluating this relation at the
onset of the instability reduces 
significantly the physically reasonable range for some parameters. 
This restriction applies only for the nematic material parameters and,
in general, nothing can be said about the corresponding smectic
parameters. We will, however, take the smectic parameters in the same
range as the nematic ones. 
If not indicated otherwise we used $L_0^{(n,s)} = 0.1$,
$L_\perp^{(n,s)} = 0.01$, $L_\parallel^{(n,s)} - L_\perp^{(n,s)} = 
0.005$, $M_0 = 10^{-4}$, $\beta_\perp^{(n,s)} = 0.01$, 
$\beta_\parallel^{(n,s)} - \beta_\perp^{(n,s)} = 0.005$, $\alpha^{(n,s)}
= 0.001$ for the plots of this section (along with parameter set
specified in the previous section).

The ansatz for $s^{(n,s)}_1$ following Eq.~(\ref{eq:ansatz}) reads
\begin{equation}
  \label{eq:s-ansatz}
  s^{(n,s)}_1 = A_s^{(n,s)} \exp[(i\omega + \frac{1}{\tau}) t]
               \sin(q_z z) \cos(q y).
\end{equation}
The modulations of $S^{(n,s)}$ in the linear analysis are
maximum at the boundaries and in phase with the layer displacement
$u$. The sign of the amplitude $A_s^{(n,s)}$ depends on the coupling
to the velocity 
field (only the anisotropic part $\beta^{(n)}_\parallel -
\beta^{(n)}_\perp$ is relevant) and on the 
coupling to the director undulations 
(via $M_{ijk}$, only for the nematic amplitudes $A_s^{(n)}$). If one
assumes that shear reduces (and does not increase) the modulus of the order
parameter, the nematic $\beta^{(n)}_\parallel - \beta^{(n)}_\perp$ is
positive
[Eqs.~(\ref{eq:s_0_n}) and (\ref{eq:s_0_n_lin})]; 
once again nothing can be
said about the smectic $\beta^{(s)}_\parallel - \beta^{(s)}_\perp$.

\begin{figure}
  \caption{\label{fig:scan-ordn}
    Out of the material parameters connected with the order parameter,
    only $\beta^{(n,s)}_\parallel - \beta^{(n,s)}_\perp$ has a
    measurable effect on the critical values. Some more parameters
    can influence the amplitudes of the order parameter undulation,
    namely $L_\perp^{(n,s)}$ and $M_0$ 
    (the latter one is only present in the case of the nematic order
    parameter). All amplitudes
    have been normalized such that $A_{\phi}= 1$. Note that the
    smectic $A_s^{(s)}$ has been multiplied by $10^3$ in the right
    column. For a better comparison we used a log-log scale in the
    lower left plot and changed the sign of $A_s^{(s)}$ in this plot.
    }
  \center\includegraphics[width=\columnwidth]{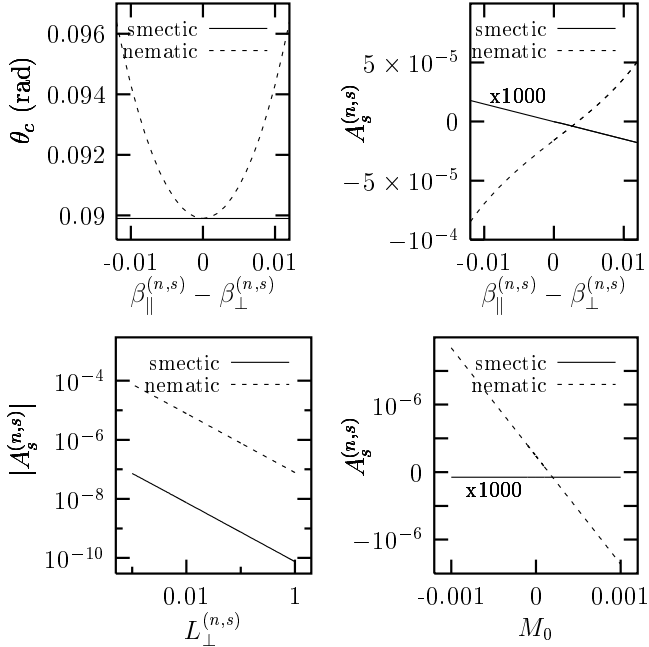}
\end{figure}

In general the critical values are not at all or only very slightly
influenced by the coupling to the modulus of the order parameter (see
Fig.~\ref{fig:scan-ordn}). Figure \ref{fig:scan-ordn} summarizes the
parameters with the largest influence on $A_s^{(n,s)}$. In almost all
investigated cases the modulation of the nematic order is much larger
than in the smectic order. Whether the order is reduced or increased in
regions where the layers are compressed depends in the
phenomenological constants $\beta^{(n,s)}_\parallel -
\beta^{(n,s)}_\perp$ and $M_0$ which have not been 
measured up to now. 

The above results reveal some interesting features. As shown in
Tab.~\ref{tab:sym}, the modulations of the order parameter change sign
under inversion of the $z$-axis. Considering the boundary condition
(i.e. taking our ansatz) this leads to the fact that the effect on the
modulus of the order parameters is maximum at the boundaries. So the
linear analysis predicts that the regions where the order parameter
is influenced most by the undulations are close to the boundaries. 
Since the probability for the formation of
defects is higher in places where the order parameter is lower, we
have identified areas where the creation of defects is facilitated. 
But
our analysis does not allow to predict the structure of the defects.
Nevertheless this effect gives a possible way how to reorient the
parallel layers. Interestingly, experiments in block copolymers 
by Laurer et al. \cite{laurer99} show a
defect structure close to the boundaries which is consistent with this
picture.

\subsection{Oscillatory instability}
\label{sec:oscill-inst} 
All our arguments in the previous sections were based on the
assumption that the undulations set in as a stationary
instability. I.e. that the oscillation rate $\omega$ in our ansatz 
Eq.~(\ref{eq:ansatz}) vanishes at onset. In this section we will
discuss the situation for non-zero $\omega$ and find that our previous
assumption was justified. In our linear analysis enters now (for the
first time in this paper) the mass density of the system, which we
will choose to be equal to unity $\rho = 1$.

\begin{figure}
  \caption[]{\label{fig:nk-oszi}
    In most parts of the scanned parameter space no possibility for an
    oscillatory instability was found. If the director field is only
    very weakly coupled to the layering (in this plot we used $B_0 =
    50$ and $\nu_i = 5$) a neutral curve for an oscillatory
    instability (\includegraphics*{line3.ps})
    appears above the stationary neutral curve
    (\includegraphics*{line1.ps}). Note that
    the critical wave vectors are close to each other for both,
    oscillatory and stationary instability. The inset shows the
    frequency along the neutral curve. 
    }
  \center\includegraphics[width=\columnwidth]{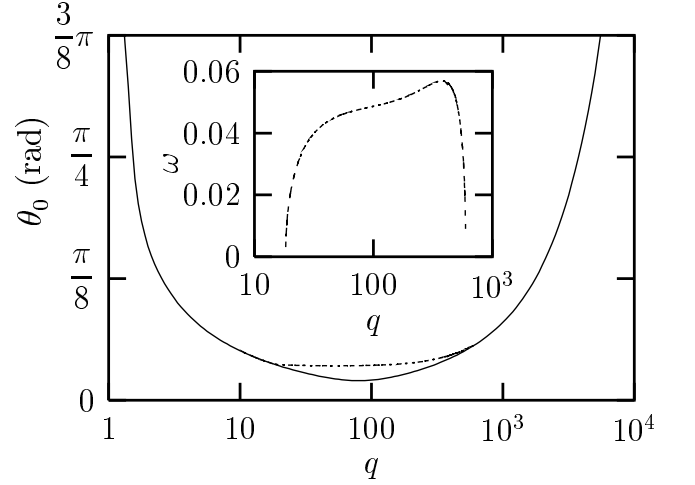}
\end{figure}

The search for a possible oscillatory instability is slightly different
from the procedure used in the stationary case. The solvability
condition of the linearized set of equations determines both the
neutral curve and the frequency along this curve. When searching for
such a solution we scanned approximately the same parameter space as
used for Figs.~\ref{fig:scan-b} -- \ref{fig:scan-lp-amp}. Since the
frequency tends to 
zero when the oscillatory neutral curve gets close to the stationary
one, we concentrated on the frequency range $0 \le \omega \le 2$ and
check in some cases for higher frequencies.

It turned out that only in cases
when the director field is very weakly coupled to the layering a
neutral curve for an oscillatory instability is possible. This weak
coupling manifests itself in small $B_1$ and $\gamma_1$, which is in
our set of dimensionless variables equivalent to large $B_0$ and
$\nu_i$. Oscillatory neutral curves were only found for $B_0 \gtrsim
100$ or $\nu_i \gtrsim 1$. 
In all investigated cases a oscillatory neutral curve is either absent
or lies above the neutral curve for a stationary instability. When a
oscillatory neutral curve is possible, it ends in the points where it
meets the stationary neutral curve (see Fig.~\ref{fig:nk-oszi}). The
corresponding frequency approaches zero in the end points of the
oscillatory neutral curve. If we ignore for the moment the stationary
neutral curve and consider only the oscillatory instability, the
corresponding critical values are found to be rather close to the
stationary one and to approach them the weaker the coupling between
the director and the layers becomes. To summarize, an oscillatory
instability was not found to be possible in all investigated cases and
seems to be extremely unlikely to occur.


\subsection{Anisotropic viscosity}
\label{sec:anisotric-viscosity}
In Fig.~\ref{fig:scan-nu} we have illustrated that a small viscosity
coefficient $\nu_2$ facilitates the onset of undulations. In this
section we will have a closer look at the effect of an anisotropic
viscosity tensor and ask whether undulations can be caused only due to
viscosity effects without any coupling to the director field (i.e. we
consider standard smectic $A$ hydrodynamics in \em this\em\;
section). 

Let us start our considerations by looking at the spatially
homogeneous state. In a sample with parallel alignment the apparent
viscosity is $\nu_3$, which can easily be seen from the force on the
upper boundary: 
\begin{equation}
  \label{eq:nu-parallel}
  \vec F_{\parallel} = \hat e_z \cdot \underline{\sigma} = 
    \dot\gamma \nu_3 \hat e_x
\end{equation}
Similarly the viscosity of a perpendicular alignment is given by
$\nu_2$:
\begin{equation}
  \label{eq:nu-perp}
  \vec F_{\perp} = \hat e_z \cdot \underline{\sigma} = 
    \dot\gamma \nu_2 \hat e_x
\end{equation}
For $\nu_2 < \nu_3$ a simple shear flow in a perpendicular
alignment causes less dissipation than in a parallel alignment. The
next step is to study the stability of these alignments in the
linear regime. Following the standard procedure (as described above)
we find a solvability condition of the linearized equations
which does not depend on the shear rate $\dot \gamma$. 
\begin{eqnarray}
  0 & = & \left\{ q^2 + 
            \lambda_p \left[
              \nu_3 \left( q^2 - q_z^2\right)^2
              + 2 \left(\nu_2 + \nu_3 \right) q^2 q_z^2
            \right]
          \right\}
          \times \nn\\&&\times 
          \left(B_0 q_z^2 + K q^4 \right) 
          \left(\nu_2 q^2 + \nu_3 q_z^2 \right)
  \label{eq:nk-visco}
\end{eqnarray}
Consequently, a parallel alignment of smectic layers is linearly stable
against undulations even if the perpendicular alignment might be more
preferable due to some thermodynamic considerations. 
As we have shown in Fig.~\ref{fig:scan-nu}, this rigorous result of
standard smectic $A$ hydrodynamics is weakened in our extended
formulation of smectic $A$ hydrodynamics. When the director can show
independent dynamics, an appropriate anisotropy of the viscosity
tensor can indeed reduce the threshold values of an undulation
instability. 

\section{Comparison to experiments and simulations}
\label{sec:outlook}
In the previous sections we have shown that the inclusion of the
director of the underlying nematic oder in the description of a
smectic $A$ like system leads to some important new features. In
general, the behavior of the director under external fields differs
from the behavior of the layer normal. In this paper we only discussed
the effect of a velocity gradient, but the effects presented here seem
to be of a more general nature and can also be applied to other fields. 
The key results of our theoretical treatment are a tilt of the
director, which is proportional to the shear rate, and an undulation
instability which sets in above a threshold value of the tilt angle
(or equivalently the shear rate). 

Both predictions are in agreement
with experimental observations. For side-chain liquid crystalline
polymers Noirez \cite{noirez00} observed a shear dependence of the
layer thickness. In the parallel orientation the layer thickness is
reduced by several percent with increasing shear. To our knowledge,
two groups have investigated the evolution of a parallel alignment to
the vesicle state for lyotropic systems (see M\"uller et
al. \cite{mueller99} and Zipfel et al. \cite{zipfel01}). In both
papers the authors argue that cylindrical structures (with an axis
along the flow direction) are observable as intermediates. 
These observed cylindrical intermediates are very close to the
undulations proposed by our theoretical treatment. 

For an approximate quantitative comparison of our theoretical results
with the experiments on lyotropic liquid crystals we make a number of 
assumptions about the material parameters. As we have shown in
Sec.~\ref{sec:stat-inst} the different approaches cause only small
variations in the critical wave number. For this estimate it suffices
to use the critical wave number obtained in our earlier work [see
Eq.~(\ref{eq:qc-old})]. For 
lyotropics it is known \cite{helfrich78,nallet89}, that the elastic
constants can be expressed as
\begin{eqnarray}
\label{eq:lyotrop-k}
K & = & \frac{\kappa}{l}
\end{eqnarray}
and
\begin{eqnarray}
\label{eq:lyotrop-b}
B = \frac{9}{64} \pi^2 \frac{(k_B T)^2}{\kappa} \; 
    \frac{l}{(l - \delta)^4},
\end{eqnarray}
where $\kappa = \alpha_{\kappa} k_B T$ is the bending modulus of a
single bilayer, $l$ is the repeat distance, $\delta$ is the membrane
thickness, $k_B$ is the Boltzmann constant, $T$ is 
temperature and $\alpha_{\kappa}$ is a dimensionless number of 
order of unity.
With this relations we can estimate the critical wave
vector for a sample of thickness $d$ using Eq.~(\ref{eq:qc-old}):
\begin{eqnarray}
\label{eq:qc-estim}
q_c^2 \approx \frac{3 \pi^2}{8 \alpha_{\kappa} d} \;
            \frac{l}{(l - \delta)^2}
\end{eqnarray}
The parameters of the experiment by Zipfel et al. \cite{zipfel01} are:
$d = 1$ mm, $\delta = 2.65$ nm, $l = 6.3$ nm and $\alpha_{\kappa} = 1.8$
\cite{zipfel01,berlepsch00}. On this basis we estimate the critical
wave length to be  of the order of 
\begin{eqnarray}
\label{eq:lambdac-estim}
\lambda_c & \approx & 6.4  \;\mu\mbox{m}
\end{eqnarray}
Zipfel et al. \cite{zipfel01} observed a vesicle radius of $3 \;
\mu$m, which is clearly compatible with our calculation. We note that
this estimate assumes that the experiments are done in the
hydrodynamic regime.

In Sect.~\ref{sec:effects-order} we have pointed out that the effect
on the order parameter is maximum close to the boundaries of the 
layer. In a reoriented sample Laurer et al. \cite{laurer99} have
identified defects near the boundary of the sample which are in
accordance with the predicted influence on the order parameter.

Molecular dynamic simulations recently made by Soddemann et
al. \cite{soddemann01} offer a very precise insight in the behavior of
the layered systems under shear. Direct comparison of these
simulations to the analytic theory presented above show a very good
agreement between both approaches \cite{soddemann02}.

The mechanism we have proposed here is somewhat similar to a shear
induced smectic-$C$ like situation. Consequently, undulations should
also be observed near the smectic-$A$--smectic-$C$ transitions. Indeed,
Johnson and Saupe \cite{johnson77} and later Kumar \cite{kumar81}
report such undulations just below the transition temperature. In the
same spirit Ribotta and Durand \cite{ribotta77} report a compression
induced smectic-$C$ like situation.

To conclude, we have shown in this work that the inclusion of nematic
degrees of freedom in the description of smectic $A$ like systems
opens the way for a shear induced destabilization of the layers under
shear. Our result are compatible with experimental observations and are
in good agreement with molecular dynamics simulations.

\acknowledgments
Partial support of this work through SFB 481 ``Komplexe
Makromolek\"ul- und Hybridsysteme in inneren und \"au{\ss}eren
Feldern'' of the Deutsche Forschungsgemeinschaft is gratefully
acknowledged.  

\begin{appendix}
\section{Minimal analytic model}
\label{sec:analyt-appr-minim}

In our earlier work \cite{auernhammer00} we considered two independent
variables (the layer displacement and the $y$-component of the
director) and found the critical values to be
\begin{eqnarray}
\label{eq:nxc-old}
  n_{x,c}^2 & = & 4 \frac{B_0}{B_0 - 2 B_1} q_z \sqrt{\frac{K}{B_0}}
	\hspace{5ex}\mbox{and} \\
\label{eq:qc-old}
  q_c^2 & = & q_z \sqrt{\frac{B_0}{K}}.
\end{eqnarray}
To compare our present analysis to these results we expand
Eqs.~(\ref{eq:min-var-f}, \ref{eq:min-var-u}) in power series in
$\theta_0$ (up to $\theta_0^2$) and take only the terms connected with
$\phi$ and $u$.  
\begin{eqnarray}
  \label{eq:app-lin-phi}
  0 & = &   A_\phi \frac{B_1}{\gamma_1} \theta_0  
	  - A_u \frac{B_1}{\gamma_1} q \\
  \label{eq:app-lin-u}
  0 & = & {} - A_u \lambda_p 
	\left[ B_1 q^2 + K q^4 + B_0 q_z^2 
	       - \frac 12 \theta_0^2 q^2 (B_0 + 2 B_1)
	\right] \nn \\ && {}
         + A_\phi \lambda_p B_1 \theta_0 q 
\end{eqnarray}
The solvability condition of Eqs.~(\ref{eq:app-lin-phi},
\ref{eq:app-lin-u}) defines the neutral curve $\theta_0(q)$ and its
minimum directly gives the critical values $\theta_c$ and $q_c$
(within the approximations of this section).
\begin{eqnarray}
  \label{eq:app-qc}
  q_c^2 & = & q_z \sqrt{\frac{B_0}{K}} \\
  \label{eq:app-theta-c}
  \theta_c^2 & = & 
    4\; q_z \;\frac{B_0}{B_0 + 2 B_1} \sqrt{\frac{K}{B_0}} \\
  \label{eq:app-gd-c}
  \dot \gamma_c & = & 
    4 \; \frac{B_1}{\gamma_1 (\lambda + 1)}
    \sqrt{q_z \;\frac{B_0}{B_0 + 2 B_1} \sqrt{\frac{K}{B_0}}}
\end{eqnarray}
The differences between Eqs.~(\ref{eq:nxc-old}, \ref{eq:qc-old}) and
(\ref{eq:app-qc} -- \ref{eq:app-gd-c}) are mainly due to the correct
normalization of $\hat p$ [see Eqs.~(\ref{eq:p_norm_1} --
\ref{eq:p_norm_3})] used in the present paper. 

To summarize, we conclude that our former results are a special case
of the present analysis when the correct normalization of $\hat p$ is
implemented. 
Especially the divergence of the critical values at $B_0 = 2 B_1$
turns out to be an artifact of the normalization of $\hat p$ used in 
Ref.\cite{auernhammer00}. 

\section{Generating the set of linear equations}
\label{sec:details-maple}

Since the theoretical methods used in this paper (irreversible
thermodynamics and linear stability analysis) offer well defined
algorithms for the generation and analysis of macroscopic hydrodynamic
equations, we performed parts of the calculation using Maple. In this
appendix we describe the key ingredients of a suitable Maple program. 
A good starting point for such an approach are the balance equations
for the unknown quantities (\ref{eq:u}, \ref{eq:kg} --
\ref{eq:order}, \ref{eq:theta}, and \ref{eq:phi}) along with the energy
density (\ref{eq:ener_coll}) in the appropriate approximation. These
equations are entered directly in Maple, with the unknown quantities
being functions of time and the spatial coordinates. The thermodynamic
forces used in these equations are determined by Eqs.~(\ref{eq:h} --
\ref{eq:xi}). For an implementation of these equations one must
take into account that Maple can only compute the derivative with
respect to constants and not with respect to functions, i.e. the
relevant functions in the energy density must be substituted
temporarily by constants. 

For the linearized set of equations we substitute the unknown
quantities by expressions of the type
\begin{equation}
  \label{eq:maple-ansatz}
  \theta(t,x,y,z) = \theta_0 + a A_\theta \sin(q_z z) \cos(q y) 
                    \exp(i \omega t)  
\end{equation}
in the governing equations. Here $a$ is a small parameter and
$A_\theta$ is the relative amplitude of the linear correction to
$\theta_0$. Expanding the substituted set of equations in a power
series in $a$ gives in zeroth order the spatially homogeneous
equations and in first order the linear set of equations, which are no
longer differential equations but algebraic ones. One obtains a matrix
representation of these 
equations by expanding them in power series of the relative amplitudes
and taking only the first order terms. After dividing by the terms
which depend on the spatial and temporal coordinates 
the Fortran code of this matrix representation is generated using the
\textsc{codegen, fortran} function of Maple and subsequently solved using
standard numerical procedures.

\end{appendix}

\bibliographystyle{apsrev}

\begin{thebibliography}{48}
\expandafter\ifx\csname natexlab\endcsname\relax\def\natexlab#1{#1}\fi
\expandafter\ifx\csname bibnamefont\endcsname\relax
  \def\bibnamefont#1{#1}\fi
\expandafter\ifx\csname bibfnamefont\endcsname\relax
  \def\bibfnamefont#1{#1}\fi
\expandafter\ifx\csname citenamefont\endcsname\relax
  \def\citenamefont#1{#1}\fi
\expandafter\ifx\csname url\endcsname\relax
  \def\url#1{\texttt{#1}}\fi
\expandafter\ifx\csname urlprefix\endcsname\relax\def\urlprefix{URL }\fi
\providecommand{\bibinfo}[2]{#2}
\providecommand{\eprint}[2][]{\url{#2}}

\bibitem[{\citenamefont{Gupta et~al.}(1996)\citenamefont{Gupta, Krishnamoorti,
  Chen, Kornfield, Smith, Satkowski, and Grothaus}}]{gupta96}
\bibinfo{author}{\bibfnamefont{V.~K.} \bibnamefont{Gupta}},
  \bibinfo{author}{\bibfnamefont{R.}~\bibnamefont{Krishnamoorti}},
  \bibinfo{author}{\bibfnamefont{Z.-R.} \bibnamefont{Chen}},
  \bibinfo{author}{\bibfnamefont{J.~A.} \bibnamefont{Kornfield}},
  \bibinfo{author}{\bibfnamefont{S.~D.} \bibnamefont{Smith}},
  \bibinfo{author}{\bibfnamefont{M.}~\bibnamefont{Satkowski}},
  \bibnamefont{and} \bibinfo{author}{\bibfnamefont{J.~T.}
  \bibnamefont{Grothaus}}, \bibinfo{journal}{Macromolecules}
  \textbf{\bibinfo{volume}{29}}, \bibinfo{pages}{875 } (\bibinfo{year}{1996}).

\bibitem[{\citenamefont{Wiesner}(1997)}]{wiesner97}
\bibinfo{author}{\bibfnamefont{U.}~\bibnamefont{Wiesner}},
  \bibinfo{journal}{Macromol. Chem. Phys.} \textbf{\bibinfo{volume}{198}},
  \bibinfo{pages}{3319 } (\bibinfo{year}{1997}).

\bibitem[{\citenamefont{Laurer et~al.}(1999)\citenamefont{Laurer, {Scott
  Pinheiro}, Polis, and Winey}}]{laurer99}
\bibinfo{author}{\bibfnamefont{J.~H.} \bibnamefont{Laurer}},
  \bibinfo{author}{\bibfnamefont{B.}~\bibnamefont{{Scott Pinheiro}}},
  \bibinfo{author}{\bibfnamefont{D.~L.} \bibnamefont{Polis}}, \bibnamefont{and}
  \bibinfo{author}{\bibfnamefont{K.~I.} \bibnamefont{Winey}},
  \bibinfo{journal}{Macromolecules} \textbf{\bibinfo{volume}{32}},
  \bibinfo{pages}{4999 } (\bibinfo{year}{1999}).

\bibitem[{\citenamefont{Zryd and Burghardt}(1998)}]{zryd98}
\bibinfo{author}{\bibfnamefont{J.~L.} \bibnamefont{Zryd}} \bibnamefont{and}
  \bibinfo{author}{\bibfnamefont{W.~R.} \bibnamefont{Burghardt}},
  \bibinfo{journal}{Macromolecules} \textbf{\bibinfo{volume}{31}},
  \bibinfo{pages}{3656 } (\bibinfo{year}{1998}).

\bibitem[{\citenamefont{Leist et~al.}(1999)\citenamefont{Leist, Maring,
  {Thurn-Albrecht}, and Wiesner}}]{leist99}
\bibinfo{author}{\bibfnamefont{H.}~\bibnamefont{Leist}},
  \bibinfo{author}{\bibfnamefont{D.}~\bibnamefont{Maring}},
  \bibinfo{author}{\bibfnamefont{T.}~\bibnamefont{{Thurn-Albrecht}}},
  \bibnamefont{and} \bibinfo{author}{\bibfnamefont{U.}~\bibnamefont{Wiesner}},
  \bibinfo{journal}{J. Chem. Phys.} \textbf{\bibinfo{volume}{110}},
  \bibinfo{pages}{8225 } (\bibinfo{year}{1999}).

\bibitem[{\citenamefont{Polis et~al.}(1999)\citenamefont{Polis, Smith, Terrill,
  Ryan, Morse, and Winey}}]{polis99}
\bibinfo{author}{\bibfnamefont{D.~L.} \bibnamefont{Polis}},
  \bibinfo{author}{\bibfnamefont{S.}~\bibnamefont{Smith}},
  \bibinfo{author}{\bibfnamefont{N.~J.} \bibnamefont{Terrill}},
  \bibinfo{author}{\bibfnamefont{A.~J.} \bibnamefont{Ryan}},
  \bibinfo{author}{\bibfnamefont{D.~C.} \bibnamefont{Morse}}, \bibnamefont{and}
  \bibinfo{author}{\bibfnamefont{K.~I.} \bibnamefont{Winey}},
  \bibinfo{journal}{Macromolecules} \textbf{\bibinfo{volume}{32}},
  \bibinfo{pages}{4668 } (\bibinfo{year}{1999}).

\bibitem[{\citenamefont{Horn and Kl{\'e}man}(1978)}]{horn78}
\bibinfo{author}{\bibfnamefont{R.~G.} \bibnamefont{Horn}} \bibnamefont{and}
  \bibinfo{author}{\bibfnamefont{M.}~\bibnamefont{Kl{\'e}man}},
  \bibinfo{journal}{Ann. Phys. (France)} \textbf{\bibinfo{volume}{3}},
  \bibinfo{pages}{229 } (\bibinfo{year}{1978}).

\bibitem[{\citenamefont{Safinya et~al.}(1991)\citenamefont{Safinya, Sirota, and
  Plano}}]{safinya91}
\bibinfo{author}{\bibfnamefont{C.~R.} \bibnamefont{Safinya}},
  \bibinfo{author}{\bibfnamefont{E.~B.} \bibnamefont{Sirota}},
  \bibnamefont{and} \bibinfo{author}{\bibfnamefont{R.~J.} \bibnamefont{Plano}},
  \bibinfo{journal}{Phys. Rev. Lett.} \textbf{\bibinfo{volume}{66}},
  \bibinfo{pages}{1986 } (\bibinfo{year}{1991}).

\bibitem[{\citenamefont{Panizza et~al.}(1995)\citenamefont{Panizza,
  Archambault, and Roux}}]{panizza95}
\bibinfo{author}{\bibfnamefont{P.}~\bibnamefont{Panizza}},
  \bibinfo{author}{\bibfnamefont{P.}~\bibnamefont{Archambault}},
  \bibnamefont{and} \bibinfo{author}{\bibfnamefont{D.}~\bibnamefont{Roux}},
  \bibinfo{journal}{J. Phys. II France} \textbf{\bibinfo{volume}{5}},
  \bibinfo{pages}{303 } (\bibinfo{year}{1995}).

\bibitem[{\citenamefont{Diat et~al.}(1993)\citenamefont{Diat, Roux, and
  Nallet}}]{diat93}
\bibinfo{author}{\bibfnamefont{O.}~\bibnamefont{Diat}},
  \bibinfo{author}{\bibfnamefont{D.}~\bibnamefont{Roux}}, \bibnamefont{and}
  \bibinfo{author}{\bibfnamefont{F.}~\bibnamefont{Nallet}},
  \bibinfo{journal}{J. Phys. II France} \textbf{\bibinfo{volume}{3}},
  \bibinfo{pages}{1427 } (\bibinfo{year}{1993}).

\bibitem[{\citenamefont{M{\"u}ller et~al.}(1999)\citenamefont{M{\"u}ller,
  B{\"o}rschig, Gronski, Schmidt, and Roux}}]{mueller99}
\bibinfo{author}{\bibfnamefont{S.}~\bibnamefont{M{\"u}ller}},
  \bibinfo{author}{\bibfnamefont{C.}~\bibnamefont{B{\"o}rschig}},
  \bibinfo{author}{\bibfnamefont{W.}~\bibnamefont{Gronski}},
  \bibinfo{author}{\bibfnamefont{C.}~\bibnamefont{Schmidt}}, \bibnamefont{and}
  \bibinfo{author}{\bibfnamefont{D.}~\bibnamefont{Roux}},
  \bibinfo{journal}{Langmuir} \textbf{\bibinfo{volume}{15}},
  \bibinfo{pages}{7558 } (\bibinfo{year}{1999}).

\bibitem[{\citenamefont{Escalante and Hoffmann}(2000)}]{escalante00}
\bibinfo{author}{\bibfnamefont{J.~I.} \bibnamefont{Escalante}}
  \bibnamefont{and} \bibinfo{author}{\bibfnamefont{H.}~\bibnamefont{Hoffmann}},
  \bibinfo{journal}{Rheol. Acta} \textbf{\bibinfo{volume}{39}},
  \bibinfo{pages}{209 } (\bibinfo{year}{2000}).

\bibitem[{\citenamefont{Zipfel et~al.}(1999)\citenamefont{Zipfel, Lindner,
  Tsianou, Alexandridis, and Richtering}}]{zipfel99b}
\bibinfo{author}{\bibfnamefont{J.}~\bibnamefont{Zipfel}},
  \bibinfo{author}{\bibfnamefont{P.}~\bibnamefont{Lindner}},
  \bibinfo{author}{\bibfnamefont{M.}~\bibnamefont{Tsianou}},
  \bibinfo{author}{\bibfnamefont{P.}~\bibnamefont{Alexandridis}},
  \bibnamefont{and}
  \bibinfo{author}{\bibfnamefont{W.}~\bibnamefont{Richtering}},
  \bibinfo{journal}{Langmuir} \textbf{\bibinfo{volume}{15}},
  \bibinfo{pages}{2599 } (\bibinfo{year}{1999}).

\bibitem[{\citenamefont{Noirez and Lapp}(1997)}]{noirez97}
\bibinfo{author}{\bibfnamefont{L.}~\bibnamefont{Noirez}} \bibnamefont{and}
  \bibinfo{author}{\bibfnamefont{A.}~\bibnamefont{Lapp}},
  \bibinfo{journal}{Phys. Rev. Lett.} \textbf{\bibinfo{volume}{78}},
  \bibinfo{pages}{70 } (\bibinfo{year}{1997}).

\bibitem[{\citenamefont{Noirez}(2000)}]{noirez00}
\bibinfo{author}{\bibfnamefont{L.}~\bibnamefont{Noirez}},
  \bibinfo{journal}{Phys. Rev. Lett.} \textbf{\bibinfo{volume}{84}},
  \bibinfo{pages}{2164 } (\bibinfo{year}{2000}).

\bibitem[{\citenamefont{Wang et~al.}(1999)\citenamefont{Wang, Newstein,
  Krishnan, Balsara, Garetz, Hammouda, and Krishnamoorti}}]{wang99}
\bibinfo{author}{\bibfnamefont{H.}~\bibnamefont{Wang}},
  \bibinfo{author}{\bibfnamefont{M.~C.} \bibnamefont{Newstein}},
  \bibinfo{author}{\bibfnamefont{A.}~\bibnamefont{Krishnan}},
  \bibinfo{author}{\bibfnamefont{N.~P.} \bibnamefont{Balsara}},
  \bibinfo{author}{\bibfnamefont{B.~A.} \bibnamefont{Garetz}},
  \bibinfo{author}{\bibfnamefont{B.}~\bibnamefont{Hammouda}}, \bibnamefont{and}
  \bibinfo{author}{\bibfnamefont{R.}~\bibnamefont{Krishnamoorti}},
  \bibinfo{journal}{Macromolecules} \textbf{\bibinfo{volume}{32}},
  \bibinfo{pages}{3695 } (\bibinfo{year}{1999}).

\bibitem[{\citenamefont{{de Gennes}}(1972)}]{degennes72}
\bibinfo{author}{\bibfnamefont{P.~G.} \bibnamefont{{de Gennes}}},
  \bibinfo{journal}{Solid State Commun.} \textbf{\bibinfo{volume}{10}},
  \bibinfo{pages}{753 } (\bibinfo{year}{1972}).

\bibitem[{\citenamefont{Martin et~al.}(1972)\citenamefont{Martin, Parodi, and
  Pershan}}]{martin72}
\bibinfo{author}{\bibfnamefont{P.~C.} \bibnamefont{Martin}},
  \bibinfo{author}{\bibfnamefont{O.}~\bibnamefont{Parodi}}, \bibnamefont{and}
  \bibinfo{author}{\bibfnamefont{P.~S.} \bibnamefont{Pershan}},
  \bibinfo{journal}{Phys. Rev. A} \textbf{\bibinfo{volume}{6}},
  \bibinfo{pages}{2401 } (\bibinfo{year}{1972}).

\bibitem[{\citenamefont{de~Gennes and Prost}(1993)}]{degennes93}
\bibinfo{author}{\bibfnamefont{P.~G.} \bibnamefont{de~Gennes}}
  \bibnamefont{and} \bibinfo{author}{\bibfnamefont{J.}~\bibnamefont{Prost}},
  \emph{\bibinfo{title}{The Physics of Liquid Crystals}}
  (\bibinfo{publisher}{Clarendon Press}, \bibinfo{address}{Oxford},
  \bibinfo{year}{1993}).

\bibitem[{\citenamefont{Pleiner and Brand}(1996)}]{pleiner96}
\bibinfo{author}{\bibfnamefont{H.}~\bibnamefont{Pleiner}} \bibnamefont{and}
  \bibinfo{author}{\bibfnamefont{H.~R.} \bibnamefont{Brand}},
  \emph{\bibinfo{title}{Pattern Formation in Liquid Crystals}}
  (\bibinfo{publisher}{Springer}, \bibinfo{address}{New York},
  \bibinfo{year}{1996}), chap. \bibinfo{chapter}{2, Hydrodynamics and
  Electrohydrodynamics of Liquid Crystals}.

\bibitem[{\citenamefont{Delaye et~al.}(1973)\citenamefont{Delaye, Ribotta, and
  Durand}}]{delaye73}
\bibinfo{author}{\bibfnamefont{M.}~\bibnamefont{Delaye}},
  \bibinfo{author}{\bibfnamefont{R.}~\bibnamefont{Ribotta}}, \bibnamefont{and}
  \bibinfo{author}{\bibfnamefont{G.}~\bibnamefont{Durand}},
  \bibinfo{journal}{Phys. Lett.} \textbf{\bibinfo{volume}{44A}},
  \bibinfo{pages}{139 } (\bibinfo{year}{1973}).

\bibitem[{\citenamefont{Clark and Meyer}(1973)}]{clark73}
\bibinfo{author}{\bibfnamefont{N.~A.} \bibnamefont{Clark}} \bibnamefont{and}
  \bibinfo{author}{\bibfnamefont{R.~B.} \bibnamefont{Meyer}},
  \bibinfo{journal}{Appl. Phys. Lett.} \textbf{\bibinfo{volume}{22}},
  \bibinfo{pages}{493 } (\bibinfo{year}{1973}).

\bibitem[{\citenamefont{Oswald and Ben-Abraham}(1982)}]{oswald82b}
\bibinfo{author}{\bibfnamefont{P.}~\bibnamefont{Oswald}} \bibnamefont{and}
  \bibinfo{author}{\bibfnamefont{S.~I.} \bibnamefont{Ben-Abraham}},
  \bibinfo{journal}{J. Phys. France} \textbf{\bibinfo{volume}{43}},
  \bibinfo{pages}{1193 } (\bibinfo{year}{1982}).

\bibitem[{\citenamefont{Wunenburger et~al.}(2000)\citenamefont{Wunenburger,
  Colin, Colin, and Roux}}]{wunenburger00}
\bibinfo{author}{\bibfnamefont{A.~S.} \bibnamefont{Wunenburger}},
  \bibinfo{author}{\bibfnamefont{A.}~\bibnamefont{Colin}},
  \bibinfo{author}{\bibfnamefont{T.}~\bibnamefont{Colin}}, \bibnamefont{and}
  \bibinfo{author}{\bibfnamefont{D.}~\bibnamefont{Roux}},
  \bibinfo{journal}{Eur. Phys. J. E} \textbf{\bibinfo{volume}{2}},
  \bibinfo{pages}{277 } (\bibinfo{year}{2000}).

\bibitem[{\citenamefont{Auernhammer
  et~al.}(2000{\natexlab{a}})\citenamefont{Auernhammer, Brand, and
  Pleiner}}]{auernhammer00}
\bibinfo{author}{\bibfnamefont{G.~K.} \bibnamefont{Auernhammer}},
  \bibinfo{author}{\bibfnamefont{H.~R.} \bibnamefont{Brand}}, \bibnamefont{and}
  \bibinfo{author}{\bibfnamefont{H.}~\bibnamefont{Pleiner}},
  \bibinfo{journal}{Rheol. Acta} \textbf{\bibinfo{volume}{39}},
  \bibinfo{pages}{215 } (\bibinfo{year}{2000}{\natexlab{a}}).

\bibitem[{\citenamefont{Auernhammer
  et~al.}(2000{\natexlab{b}})\citenamefont{Auernhammer, Brand, and
  Pleiner}}]{auernhammer00a}
\bibinfo{author}{\bibfnamefont{G.~K.} \bibnamefont{Auernhammer}},
  \bibinfo{author}{\bibfnamefont{H.~R.} \bibnamefont{Brand}}, \bibnamefont{and}
  \bibinfo{author}{\bibfnamefont{H.}~\bibnamefont{Pleiner}},
  \bibinfo{journal}{Proceedings Freiburger Arbeitstagung Fl{\"u}ssigkristalle}
  \textbf{\bibinfo{volume}{29}}, \bibinfo{pages}{V19}
  (\bibinfo{year}{2000}{\natexlab{b}}).

\bibitem[{\citenamefont{Bruinsma and Rabin}(1992)}]{bruinsma92}
\bibinfo{author}{\bibfnamefont{R.}~\bibnamefont{Bruinsma}} \bibnamefont{and}
  \bibinfo{author}{\bibfnamefont{Y.}~\bibnamefont{Rabin}},
  \bibinfo{journal}{Phys. Rev. A} \textbf{\bibinfo{volume}{45}},
  \bibinfo{pages}{994 } (\bibinfo{year}{1992}).

\bibitem[{\citenamefont{Zilman and Granek}(1999)}]{zilman99}
\bibinfo{author}{\bibfnamefont{A.~G.} \bibnamefont{Zilman}} \bibnamefont{and}
  \bibinfo{author}{\bibfnamefont{R.}~\bibnamefont{Granek}},
  \bibinfo{journal}{Eur. Phys. J. B} \textbf{\bibinfo{volume}{11}},
  \bibinfo{pages}{593 } (\bibinfo{year}{1999}).

\bibitem[{\citenamefont{Williams and MacKintosh}(1994)}]{williams94}
\bibinfo{author}{\bibfnamefont{D.~R.~M.} \bibnamefont{Williams}}
  \bibnamefont{and} \bibinfo{author}{\bibfnamefont{F.~C.}
  \bibnamefont{MacKintosh}}, \bibinfo{journal}{Macromolecules}
  \textbf{\bibinfo{volume}{27}}, \bibinfo{pages}{7677 } (\bibinfo{year}{1994}).

\bibitem[{\citenamefont{{de Gennes}}(1973)}]{degennes73}
\bibinfo{author}{\bibfnamefont{P.~G.} \bibnamefont{{de Gennes}}},
  \bibinfo{journal}{Mol. Cryst. Liq. Cryst.} \textbf{\bibinfo{volume}{21}},
  \bibinfo{pages}{49 } (\bibinfo{year}{1973}).

\bibitem[{\citenamefont{Litster et~al.}(1979)\citenamefont{Litster,
  Als-Nielson, Birgeneau, Dana, Davidov, Garcia-Golding, Kaplan, Safinya, and
  Schaetzing}}]{litster79}
\bibinfo{author}{\bibfnamefont{J.~D.} \bibnamefont{Litster}},
  \bibinfo{author}{\bibfnamefont{J.}~\bibnamefont{Als-Nielson}},
  \bibinfo{author}{\bibfnamefont{R.~J.} \bibnamefont{Birgeneau}},
  \bibinfo{author}{\bibfnamefont{S.~S.} \bibnamefont{Dana}},
  \bibinfo{author}{\bibfnamefont{D.}~\bibnamefont{Davidov}},
  \bibinfo{author}{\bibfnamefont{F.}~\bibnamefont{Garcia-Golding}},
  \bibinfo{author}{\bibfnamefont{M.}~\bibnamefont{Kaplan}},
  \bibinfo{author}{\bibfnamefont{C.~R.} \bibnamefont{Safinya}},
  \bibnamefont{and}
  \bibinfo{author}{\bibfnamefont{R.}~\bibnamefont{Schaetzing}},
  \bibinfo{journal}{J. Phys. Coll. C3} \textbf{\bibinfo{volume}{40}},
  \bibinfo{pages}{C3} (\bibinfo{year}{1979}).

\bibitem[{\citenamefont{Garoff and Meyer}(1977)}]{garoff77}
\bibinfo{author}{\bibfnamefont{S.}~\bibnamefont{Garoff}} \bibnamefont{and}
  \bibinfo{author}{\bibfnamefont{R.~B.} \bibnamefont{Meyer}},
  \bibinfo{journal}{Phys. Rev. Lett.} \textbf{\bibinfo{volume}{38}},
  \bibinfo{pages}{848 } (\bibinfo{year}{1977}).

\bibitem[{\citenamefont{de~Groot and Mazur}(1962)}]{degroot62}
\bibinfo{author}{\bibfnamefont{S.~R.} \bibnamefont{de~Groot}} \bibnamefont{and}
  \bibinfo{author}{\bibfnamefont{P.}~\bibnamefont{Mazur}},
  \emph{\bibinfo{title}{Non-equilibrium Thermodynamics}}
  (\bibinfo{publisher}{North-Holland}, \bibinfo{address}{Amsterdam},
  \bibinfo{year}{1962}).

\bibitem[{\citenamefont{Liu}(1979)}]{liu79}
\bibinfo{author}{\bibfnamefont{M.}~\bibnamefont{Liu}}, \bibinfo{journal}{Phys.
  Rev. A} \textbf{\bibinfo{volume}{19}}, \bibinfo{pages}{2090 }
  (\bibinfo{year}{1979}).

\bibitem[{\citenamefont{Brand and Kawasaki}(1986)}]{brand86a}
\bibinfo{author}{\bibfnamefont{H.~R.} \bibnamefont{Brand}} \bibnamefont{and}
  \bibinfo{author}{\bibfnamefont{K.}~\bibnamefont{Kawasaki}},
  \bibinfo{journal}{J. Phys. C} \textbf{\bibinfo{volume}{19}},
  \bibinfo{pages}{937 } (\bibinfo{year}{1986}).

\bibitem[{\citenamefont{Marignan et~al.}(1983)\citenamefont{Marignan, O.Parodi,
  and {Dubois-Violette}}}]{marignan83}
\bibinfo{author}{\bibfnamefont{J.}~\bibnamefont{Marignan}},
  \bibinfo{author}{\bibnamefont{O.Parodi}}, \bibnamefont{and}
  \bibinfo{author}{\bibfnamefont{E.}~\bibnamefont{{Dubois-Violette}}},
  \textbf{\bibinfo{volume}{44}}, \bibinfo{pages}{263 } (\bibinfo{year}{1983}).

\bibitem[{\citenamefont{Pleiner and Brand}(1999)}]{pleiner99}
\bibinfo{author}{\bibfnamefont{H.}~\bibnamefont{Pleiner}} \bibnamefont{and}
  \bibinfo{author}{\bibfnamefont{H.~R.} \bibnamefont{Brand}},
  \bibinfo{journal}{Physica A} \textbf{\bibinfo{volume}{265}},
  \bibinfo{pages}{62 } (\bibinfo{year}{1999}).

\bibitem[{\citenamefont{Forster et~al.}(1971)\citenamefont{Forster, Lubensky,
  Martin, Swift, and Pershan}}]{forster71}
\bibinfo{author}{\bibfnamefont{D.}~\bibnamefont{Forster}},
  \bibinfo{author}{\bibfnamefont{T.~C.} \bibnamefont{Lubensky}},
  \bibinfo{author}{\bibfnamefont{P.~C.} \bibnamefont{Martin}},
  \bibinfo{author}{\bibfnamefont{J.}~\bibnamefont{Swift}}, \bibnamefont{and}
  \bibinfo{author}{\bibfnamefont{P.~S.} \bibnamefont{Pershan}},
  \bibinfo{journal}{Phys. Rev. Lett.} \textbf{\bibinfo{volume}{26}},
  \bibinfo{pages}{1016 } (\bibinfo{year}{1971}).

\bibitem[{\citenamefont{Graf et~al.}(1992)\citenamefont{Graf, Kneppe, and
  Schneider}}]{graf92}
\bibinfo{author}{\bibfnamefont{H.-H.} \bibnamefont{Graf}},
  \bibinfo{author}{\bibfnamefont{H.}~\bibnamefont{Kneppe}}, \bibnamefont{and}
  \bibinfo{author}{\bibfnamefont{F.}~\bibnamefont{Schneider}},
  \bibinfo{journal}{Mol. Phys.} \textbf{\bibinfo{volume}{77}},
  \bibinfo{pages}{521 } (\bibinfo{year}{1992}).

\bibitem[{\citenamefont{Soddemann et~al.}(to be
  published)\citenamefont{Soddemann, Auernhammer, Guo, D{\"u}nweg, and
  Kremer}}]{soddemann02}
\bibinfo{author}{\bibfnamefont{T.}~\bibnamefont{Soddemann}},
  \bibinfo{author}{\bibfnamefont{G.~K.} \bibnamefont{Auernhammer}},
  \bibinfo{author}{\bibfnamefont{H.~X.} \bibnamefont{Guo}},
  \bibinfo{author}{\bibfnamefont{B.}~\bibnamefont{D{\"u}nweg}},
  \bibnamefont{and} \bibinfo{author}{\bibfnamefont{K.}~\bibnamefont{Kremer}}
  (\bibinfo{year}{to be published}).

\bibitem[{\citenamefont{Zipfel et~al.}(2001)\citenamefont{Zipfel, Nettesheim,
  Linder, Le, Olsson, and Richtering}}]{zipfel01}
\bibinfo{author}{\bibfnamefont{J.}~\bibnamefont{Zipfel}},
  \bibinfo{author}{\bibfnamefont{F.}~\bibnamefont{Nettesheim}},
  \bibinfo{author}{\bibfnamefont{P.}~\bibnamefont{Linder}},
  \bibinfo{author}{\bibfnamefont{T.~D.} \bibnamefont{Le}},
  \bibinfo{author}{\bibfnamefont{U.}~\bibnamefont{Olsson}}, \bibnamefont{and}
  \bibinfo{author}{\bibfnamefont{W.}~\bibnamefont{Richtering}},
  \bibinfo{journal}{Europhys. Lett.} \textbf{\bibinfo{volume}{53}},
  \bibinfo{pages}{335 } (\bibinfo{year}{2001}).

\bibitem[{\citenamefont{Helfrich}(1978)}]{helfrich78}
\bibinfo{author}{\bibfnamefont{W.}~\bibnamefont{Helfrich}},
  \bibinfo{journal}{Z. Naturforsch.} \textbf{\bibinfo{volume}{33a}},
  \bibinfo{pages}{305 } (\bibinfo{year}{1978}).

\bibitem[{\citenamefont{Nallet et~al.}(1989)\citenamefont{Nallet, Roux, and
  Prost}}]{nallet89}
\bibinfo{author}{\bibfnamefont{F.}~\bibnamefont{Nallet}},
  \bibinfo{author}{\bibfnamefont{D.}~\bibnamefont{Roux}}, \bibnamefont{and}
  \bibinfo{author}{\bibfnamefont{J.}~\bibnamefont{Prost}}, \bibinfo{journal}{J.
  Phys. France} \textbf{\bibinfo{volume}{50}}, \bibinfo{pages}{3147 }
  (\bibinfo{year}{1989}).

\bibitem[{\citenamefont{{von Berlepsch} and {de Vries}}(2000)}]{berlepsch00}
\bibinfo{author}{\bibfnamefont{H.}~\bibnamefont{{von Berlepsch}}}
  \bibnamefont{and} \bibinfo{author}{\bibfnamefont{R.}~\bibnamefont{{de
  Vries}}}, \bibinfo{journal}{Eur. Phys. J. E} \textbf{\bibinfo{volume}{1}},
  \bibinfo{pages}{141 } (\bibinfo{year}{2000}).

\bibitem[{\citenamefont{Soddemann et~al.}(2001)\citenamefont{Soddemann,
  D{\"u}nweg, and Kremer}}]{soddemann01}
\bibinfo{author}{\bibfnamefont{T.}~\bibnamefont{Soddemann}},
  \bibinfo{author}{\bibfnamefont{B.}~\bibnamefont{D{\"u}nweg}},
  \bibnamefont{and} \bibinfo{author}{\bibfnamefont{K.}~\bibnamefont{Kremer}},
  \bibinfo{journal}{Eur. Phys. J. E} \textbf{\bibinfo{volume}{6}},
  \bibinfo{pages}{409 } (\bibinfo{year}{2001}).

\bibitem[{\citenamefont{Johnson and Saupe}(1977)}]{johnson77}
\bibinfo{author}{\bibfnamefont{D.}~\bibnamefont{Johnson}} \bibnamefont{and}
  \bibinfo{author}{\bibfnamefont{A.}~\bibnamefont{Saupe}},
  \bibinfo{journal}{Phys. Rev. A} \textbf{\bibinfo{volume}{15}},
  \bibinfo{pages}{2079 } (\bibinfo{year}{1977}).

\bibitem[{\citenamefont{Kumar}(1981)}]{kumar81}
\bibinfo{author}{\bibfnamefont{S.}~\bibnamefont{Kumar}},
  \bibinfo{journal}{Phys. Rev. A} \textbf{\bibinfo{volume}{23}},
  \bibinfo{pages}{3207 } (\bibinfo{year}{1981}).

\bibitem[{\citenamefont{Ribotta and Durand}(1977)}]{ribotta77}
\bibinfo{author}{\bibfnamefont{R.}~\bibnamefont{Ribotta}} \bibnamefont{and}
  \bibinfo{author}{\bibfnamefont{G.}~\bibnamefont{Durand}},
  \bibinfo{journal}{J. Phys. France} \textbf{\bibinfo{volume}{38}},
  \bibinfo{pages}{179 } (\bibinfo{year}{1977}).

\end{thebibliography}

\end{document}